\documentclass[aps,prb,twocolumn,groupedaddress,showpacs]{revtex4-1}
\pdfoutput=1

\usepackage{physics}
\usepackage{graphicx}
\usepackage{hyperref}
\usepackage[caption=false]{subfig}

\begin{document}

\title{Compression of Correlation Matrices and an Efficient Method for Forming 
Matrix Product States of Fermionic Gaussian States}

\author{Matthew T. Fishman}
\affiliation{Institute for Quantum Information and Matter, 
California Institute of Technology, Pasadena, CA 91125, USA}

\author{Steven R. White}
\affiliation{Department of Physics and Astronomy, 
University of California, Irvine, CA 92697, USA}

\date{\today}

\begin{abstract}
Here we present an efficient and numerically stable procedure for compressing a 
correlation matrix into a set of local unitary single-particle gates, which leads to a 
very efficient way of forming the matrix product state (MPS) approximation of a pure 
fermionic Gaussian state, such as the ground state of a quadratic Hamiltonian. 
The procedure involves successively diagonalizing subblocks of the correlation matrix to 
isolate local states which are purely occupied or unoccupied. 
A small number of nearest neighbor unitary gates isolates each local state. 
The MPS of this state is formed by applying the many-body version of these gates to a 
product state.  
We treat the simple case of compressing the correlation matrix of spinless free fermions with
definite particle number in detail, though the procedure is easily extended to fermions 
with spin and more general BCS states (utilizing the formalism of Majorana modes).
We also present a DMRG-like algorithm to obtain the compressed correlation matrix directly 
from a hopping Hamiltonian.
In addition, we discuss a slight variation of the procedure which leads to a simple 
construction of the multiscale entanglement renormalization ansatz (MERA) of a fermionic 
Gaussian state, and present a simple picture of orthogonal wavelet transforms in terms of the
gate structure we present in this paper.
As a simple demonstration we analyze the Su-Schrieffer-Heeger model (free fermions on a 1D 
lattice with staggered hopping amplitudes). 
\end{abstract}


\maketitle

\section{Introduction}

One of the strengths of the density matrix renormalization group (DMRG)
\cite{white1992,white1993}, and tensor network states in general, is that their power to 
simulate strongly correlated systems does not require the interactions to be weak.  In 
fact, in fermion systems such as the Hubbard model, DMRG is {\it more} accurate for 
larger interactions.  The matrix product state (MPS) representation of the wavefunction, 
which DMRG implicitly uses, more efficiently compresses the wavefunction when interactions
are strong, due to lower entanglement in a real-space basis.  

In this paper, we introduce a new algorithm for efficiently producing an MPS 
representation for ground states of {\it noninteracting}  fermion systems.  Why is this 
useful, when DMRG is most useful in the opposite regime? This would be a valuable tool 
in a number of situations.  For example, a powerful and widely used class of variational 
wavefunctions for strongly interacting systems begin with a mean-field fermionic 
wavefunction, and then one applies a Gutzwiller projection to reduce or eliminate double 
occupancy.\cite{gutzwiller1963}  It could be very useful to find the overlaps of a DMRG 
ground state with a variety of such Gutzwiller states to help understand and classify the 
ground state. Once one has the MPS representation of the mean field state, the
Gutzwiller projection is very easy, fast, and exact, whereas in other approaches it
usually must be implemented with Monte Carlo. One might also begin a DMRG simulation with 
such a variational state, or in some cases with a mean field state without the Gutzwiller 
projection. Being able to represent fermion determinantal states as MPS's in a very
efficient way also opens the door to using DMRG ground states as minus-sign constraints 
in determinantal quantum Monte Carlo, in particular in Zhang's constrained path Monte 
Carlo (CPMC) method\cite{zhang1995,zhang1997}.  In this case one would hope that, for 
systems too big for accurate DMRG, at least the qualitative structure of the ground state 
could be captured by DMRG, and then the results could be made quantitative with the Monte 
Carlo method.

The basis of our approach shares ideas with DMRG. Matrix product state representations 
exploit a property of the state (low entanglement) to compress the information in the 
state. Fermionic Gaussian states (the general class of states which includes both fermion
determinants, BCS states, and free fermion thermal states) are also compressible, as we will 
show. The properties of a Gaussian state are completely defined by its correlation
matrix.  For the case of a fermion determinant, the correlation matrix has eigenvalues
which are either 0 or 1, i.e. they carry only a limited amount of information,
indicating that the state can be compressed.  In particular, one can perform an
arbitrary single-particle change of basis within the occupied states, or
within the unoccupied states, without changing the determinantal state. Tensor network
methods in the context of fermionic Gaussian states have been studied previously in the 
context of the multiscale entanglement renormalization ansatz (MERA)\cite{evenbly2010-1} 
and projected entangled pair states (PEPS)\cite{kraus2010}, however here we present a simple
and easily generalizable formalism and construction starting with an efficient method for
forming the MPS of a fermionic Gaussian state. We also present a new and 
simpler method for obtaining a fermionic Gaussian MERA (GMERA), the MERA of a fermionic 
Gaussian state, as a simple extension.

Our approach to producing the MPS of a fermionic Gaussian state also produces a compressed 
form of the correlation matrix itself, which we call a fermionic Gaussian MPS (GMPS), which 
might be useful in very different contexts where the single-particle matrices are very large. 
This compressed form expresses the $N\times N$ correlation matrix in terms of $O(BN)$ real 
angles which parametrize nearest neighbor rotation gates, where $B\ll N$ for states with low 
entanglement.  The compressed form can be utilized directly. For example, ordinarily 
multiplying an arbitrary vector by the correlation matrix, which is not sparse, requires 
$O(N^2)$ operations, but by using the compressed form only $O(BN)$ operations are needed.
For simplicity, the algorithm we introduce first utilizes the correlation matrix as the 
initial input. However, in Appendix~\ref{appendix2} we present a DMRG algorithm in the 
single particle context, which we call fermionic Gaussian DMRG (GDMRG), that starts with a 
single particle Hamiltonian matrix and outputs the ground state correlation matrix in 
compressed form as a GMPS at a greatly reduced cost compared to directly diagonalizing the 
Hamiltonian matrix, $O(B^3 N)$ as opposed to $O(N^3)$. This algorithm exploits the close 
relationship between the correlation matrix and the density matrix of a many particle 
state, and many tensor network algorithms can similarly be translated into a single 
particle framework. 

The paper is organized as follows. Section~\ref{background} gives a brief overview of 
fermionic Gaussian states and correlation matrices, including an introduction to the 
entanglement of these states. 
In Section~\ref{algorithms}, we give detailed descriptions of the new algorithms.
Section~\ref{cor_to_gmps} covers our new procedure for compressing a correlation matrix as 
a GMPS.
Section~\ref{cor_to_gmera} presents a variation of this method to obtain a GMERA. 
In Section~\ref{gmera_wavelet} we give a brief introduction to how the GMERA gate structure 
relates to wavelet transforms.
Section~\ref{gmps_to_mps} covers the procedure for turning the gates obtained from 
compressing the correlation matrix into a many-body MPS approximation of the Gaussian state.
Finally, Section~\ref{results} shows numerical results for the algorithms covered in the 
paper.

\section{Background on Fermionic Gaussian States and Correlation Matrices}
\label{background}

Consider the Hamiltonian for a 1D system of noninteracting fermions
\begin{eqnarray}
\hat{H} = \sum_{i,j=1}^N \hat{a}_i^{\dagger} H_{ij} \hat{a}_j
\label{eq:numconsham}
\end{eqnarray}
where $a_i$ and $a_i^\dagger$ are fermion operators and
$H=[H_{ij}]$ is a symmetric matrix ($H = H^{\dagger}$). We assume that the 
Hamiltonian terms are local (so the matrix $H$ is band-diagonal).

Diagonalizing the matrix $H$, we have $H = U D U^{\dagger}$ where $U$ 
is orthogonal and $D$ is diagonal such that $D_{kk'} = \epsilon_k \delta_{kk'}$.
The Hamiltonian can then be put into diagonal form,
\begin{eqnarray}
\hat{H} = \sum_{k=1}^N \epsilon_k \hat{a}_k^{\dagger} \hat{a}_k
\end{eqnarray}
where the operators which create the single particle energy eigenstates are 
\begin{eqnarray}
\hat{a}^{\dagger}_k = 
\sum_{i=1}^N U_{ik} \hat{a}^{\dagger}_i .
\label{eq:primebasis}
\end{eqnarray}
Assuming $\epsilon_k \le \epsilon_{k'}$ if $k<k'$,
the ground state is
\begin{eqnarray}
\ket{\psi_0} = \prod_{k=1}^{N_F} \hat{a}_k^{\dagger} \ket{\Omega}.
\label{eq:numconsgs}
\end{eqnarray}
where $N_F$ is the number of particles in the system.

The \emph{correlation matrix} is
\begin{eqnarray}
\Lambda_{ij} = \left< \hat{a}^{\dagger}_i \hat{a}_j \right> = 
    \sum_{k=1}^{N_F} U^*_{ik} U_{jk}.
\end{eqnarray}
The correlation matrix fully characterizes $\ket{\psi_0}$ because all correlation 
functions, and therefore all observables, can be factorized into two-point correlators 
using Wick's theorem. Note that the eigenstates of $H$ are also the eigenstates of 
$\Lambda$ (the same $U$ that diagonalizes $H$ also diagonalizes $\Lambda$). However, the 
eigenvalues of $\Lambda$ are either 1 (occupied) or 0 (unoccupied).
The massive degeneracy of $\Lambda$ means that we can make arbitrary changes of basis 
among the eigenstates of $\Lambda$ as long as we do not mix occupied and unoccupied 
states. 

\begin{figure}
    \subfloat[Eigenvalues and entanglement entropy.]{
      \includegraphics[width=0.5\textwidth]{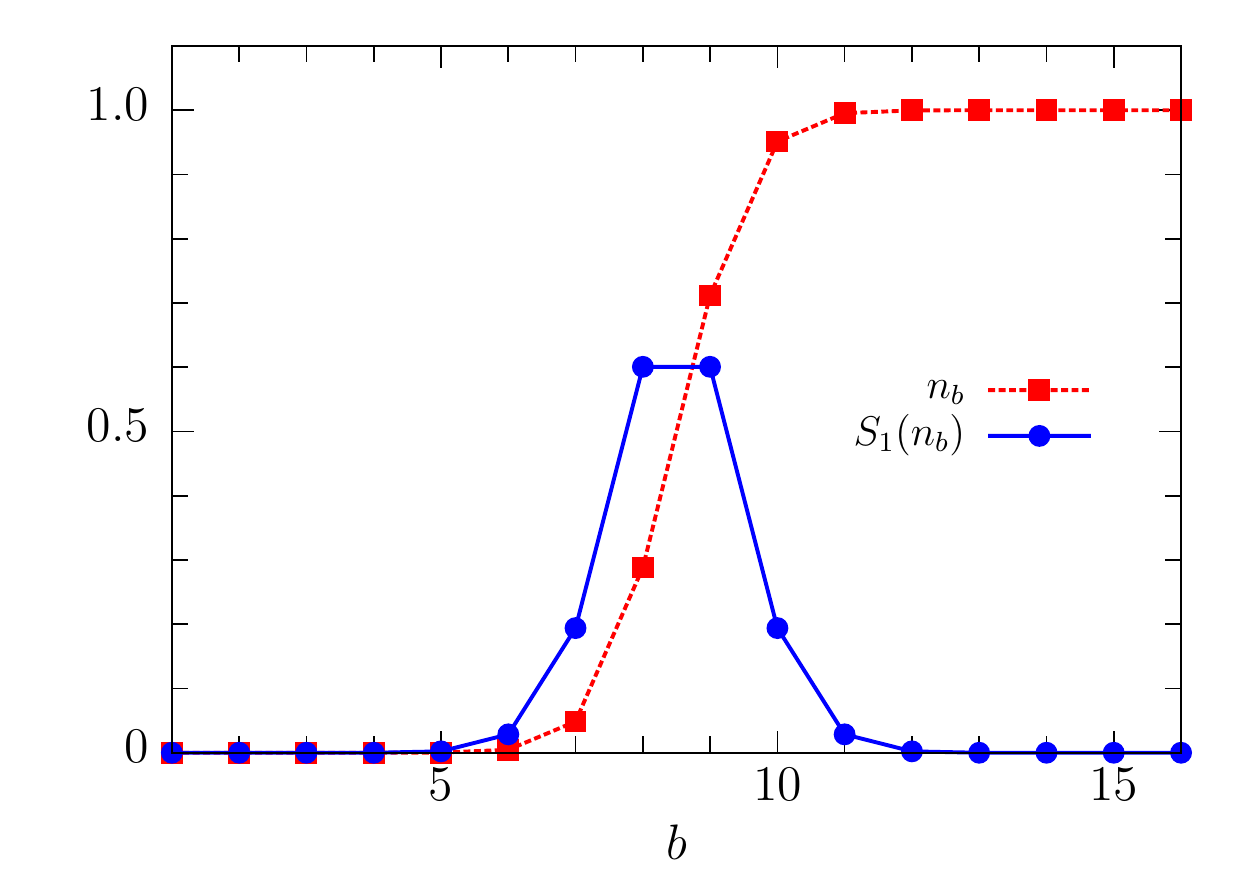}%
      } \\
    \subfloat[Example eigenvectors.]{
      \includegraphics[width=0.5\textwidth]{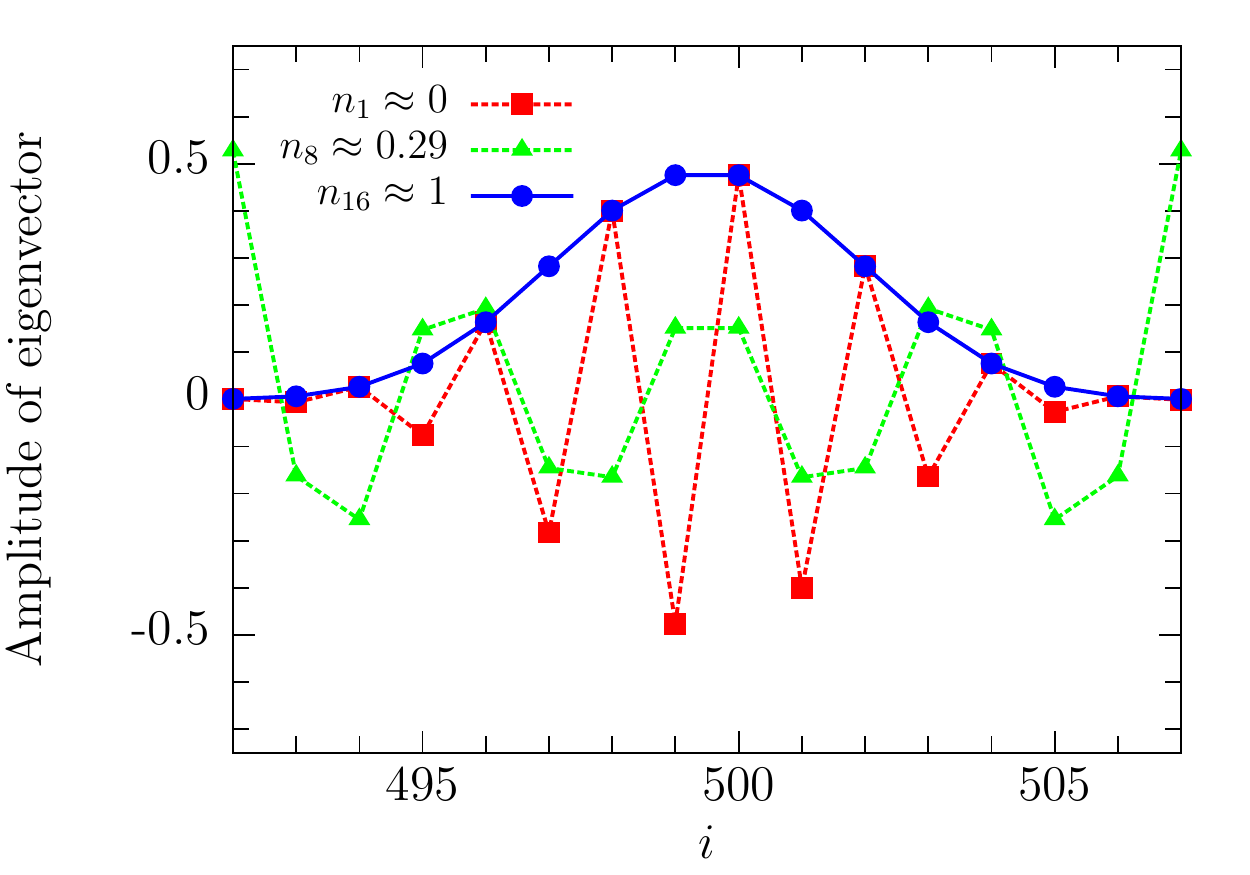}%
      }
    \caption{Fig.~\ref{entanglement}(a) shows the occupations $n_b$ and corresponding
        entanglement $S_1(n_b)$ from diagonalizing a block of $B=16$
        sites in the middle of a system of free gapless fermions on $N=1000$ sites at half
        filling. The minimum and maximum eigenvalues, $n_1$ and $n_{16}$, differ from
        0 and 1 by $\approx 1.74\times 10^{-11}$. The eigenvalues closest to 1/2, 
        $1/2-n_8=n_9-1/2\approx 0.21$, have entropies $S_1(n_8)=S_1(n_9)\approx 0.60$, 
        which are close to the maximum of $S_1(1/2)=\log(2)\approx 0.69$.
        Fig.~\ref{mode}(b) shows examples of eigenvectors from the same diagonalization.
        The eigenvectors with eigenvalues near 0 and 1, which contribute very little
        to the entanglement, are localized in the middle of the block, while the 
        eigenvectors with eigenvalues closer to 1/2 which contribute most to the 
        entanglement have large support on the edges of the block.}
    \label{entanglement}
\end{figure}

In our procedure, we will be interested in finding localized eigenvectors of the correlation
matrix which are (approximately) fully occupied or unoccupied. 
By rotating into the basis of these eigenvectors, we can locally diagonalize the 
correlation matrix, which will lead to a compression of the state.
These eigenvectors have eigenvalues near 1 or 0, which makes them (approximate) eigenvectors 
of the entire correlation matrix and therefore uncorrelated with the rest of the system. 
What makes it possible to find a localized eigenvector?

The answer is the limited entanglement structure of the states we are interested in
(ground states of local Hamiltonians).
Consider the entanglement entropy of our fermionic Gaussian state, which can be 
calculated directly from the correlation matrix. 
Divide the system into an arbitrary subblock $\mathcal{B}$ of $B$ sites (with the 
corresponding submatrix of $\Lambda$, which we call $\Lambda_{\mathcal{B}}$) and the 
rest of the system. 
We would like to know how large of a block size $B$ we need to find a localized eigenvector.
If the matrix $\Lambda_{\mathcal{B}}$ has eigenvalues $\{n_b\}$ for 
$b\in\mathcal{B}$, with $0 \le n_b \le 1$, 
the entanglement entropy of the subblock $\mathcal{B}$, 
$S_B\equiv -\Tr[\hat{\rho}_{\mathcal{B}}\log(\hat{\rho}_{\mathcal{B}})]$ 
(where $\hat{\rho}_{\mathcal{B}}$ is the reduced density matrix of the state in 
subblock $\mathcal{B}$), is 
\begin{eqnarray}
S_B(\{n_b\})=\sum_{b\in\mathcal{B}} S_1(n_b)
\label{entropy}
\end{eqnarray}
where $S_1(n_b)=-[n_b\log(n_b)+(1-n_b)\log(1-n_b)]$.
This expression has been shown elsewhere\cite{latorre2009,peschel2009,vidal2003,latorre2003}.
We show a simple, self contained derivation of it in Appendix~\ref{appendix1}.
Note that $S_1(n_b)$ vanishes for both $n_b \to 0$ and $n_b \to 1$.

The maximum amount of entanglement a block of size $B$ can contain is when $n_b=1/2$ for all 
$b\in\mathcal{B}$, so $S_B\leq B\log(2)$.
This reflects a volume law entanglement in the ``volume" B.  However, ground states of 1D 
local Hamiltonians have entanglement that is much smaller, either of order unity (if the 
system is gapped), or the entanglement grows as $\log(B)$ if the system is gapless.  
To avoid the volume entanglement, most of the block eigenvalues $n_b$ must be exponentially 
close to 0 or 1.
In other words, as soon as we make $B$ big enough so that the entanglement begins to saturate,
except for a possible slow logarithmic growth, we should find at least one eigenvalue very 
close to 0 or 1.
For gapless free fermions in 1D on $N=1000$ sites, we show example eigenvalues,
eigenvectors, and corresponding entanglements of a block of $B=16$ in the
middle of the correlation matrix in Fig.~\ref{entanglement}. Even for gapless
free Fermions, with a block size of only $B=16$ we find many eigenvalues near 0
or 1 (many localized eigenvectors).  We use this observation next to develop
algorithms to locally diagonalize correlation matrices and in the process find
a very compressed form.

\section{Algorithms}
\label{algorithms}

\subsection{Compressing a Correlation Matrix as a GMPS}
\label{cor_to_gmps}

We begin the procedure by diagonalizing the upper left $B \times B$ subblock of
a correlation matrix $\Lambda$ of a pure fermionic Gaussian state.
Assume that the state has some local entanglement structure, for example it is the ground
state of a local Hamiltonian in 1D.
For now, we imagine our system has open boundary conditions.
Let $\mathcal{B}$ be the group of sites $1, \ldots B$ on the left end of the system, and 
$\Lambda_{\mathcal{B}}$ be the associated subblock of $\Lambda$.  
Also, let $\{n_b\}$ be the eigenvalues of $\Lambda_{\mathcal{B}}$ for $b\in\mathcal{B}$ where 
$0 \leq n_b \leq 1$. (This constraint on the eigenvalues of the 
subblock follows from the fact that both $\Lambda$ and $1-\Lambda$ are positive 
semi-definite.) 

\begin{figure}
\includegraphics[scale=0.7]{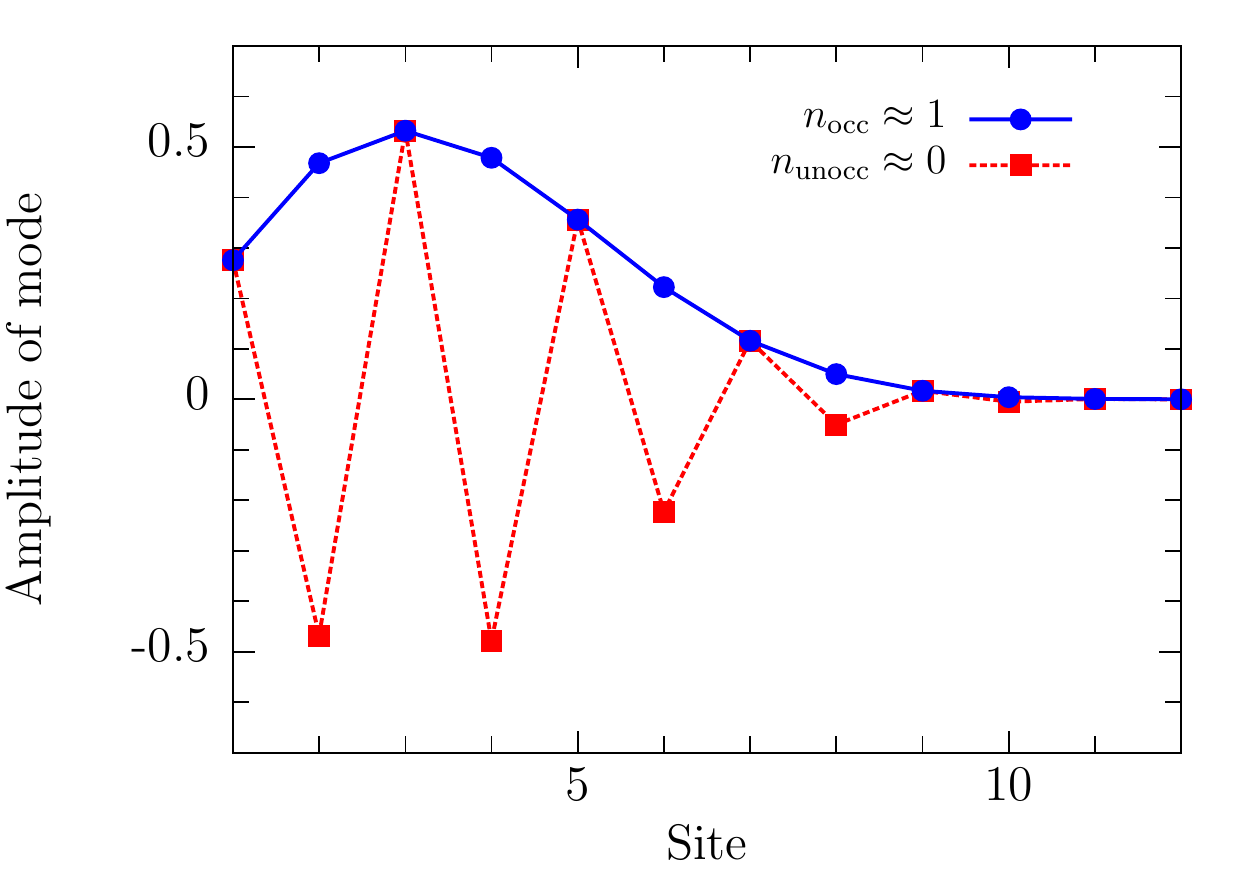}
\caption[]{Examples of approximate occupied and unoccupied eigenvectors of $\Lambda$ 
    obtained from diagonalizing $\Lambda_{\mathcal{B}}$ where subblock $\mathcal{B}$
    are sites $1,\ldots,B$. $\Lambda$ is formed from the ground state of 
    $\hat{H}=-t\sum_{i=1}^{N-1} (\hat{a}^{\dagger}_i\hat{a}_{i+1}+h.c.)$ for $N=1024$ 
    at half filling ($N_F=N/2$). A block size of $B=12$ is used. Eigenvectors
    with highest ($n_{\text{occ}}$) and lowest ($n_{\text{unocc}}$) eigenvalues found
    from diagonalizing subblock $\mathcal{B}$ are shown. We find 
    $1-n_{\text{occ}}=2.4\times 10^{-15}$ and 
    $n_{\text{unocc}}=7.3\times 10^{-16}$, so the occupations are accurate to 
    nearly machine double precision. $1-n_{\text{occ}}$ and $n_{\text{unocc}}$ should
    be equal at half filling (because of particle-hole symmetry), but are different 
    in this case as a result of roundoff errors.}
\label{left_mode}
\end{figure}

We increase $B$ until we find some $n_b$ that is nearly 1 or 0 within a specified tolerance, 
e.g. $10^{-6}$. 
The closer the eigenvalue is to 1 or 0, the more accurate the compression, but a larger 
block size translates to more gates and a bigger bond dimension of the MPS we will form. 
In Fig.~\ref{left_mode} we show the most occupied and unoccupied eigenvectors of 
$\Lambda_{\mathcal{B}}$ for $B=12$ for a system of gapless free fermions in 1D with 
$N=1024$ sites. 
We see that $B=12$ is sufficient to give deviations from occupancies of 0 or 1 to nearly 
machine double precision. 
The eigenvalues found in the bulk likely will not be as accurate, because states in the bulk
will generally be more entangled than the ones on the edge.
The smooth fall-off to zero at the right edge of the block is characteristic of these modes 
and is a result of diagonalizing the block on the left-most boundary of the system. 
The localized states we find here are {\it least} entangled with the rest of the system.
This is in contrast to the dominant Schmidt states that are utilized within DMRG which have 
degrees of freedom that are localized at the edge of the block. 

The eigenvector $\vec{v}$ which is least entangled is also 
an approximate eigenvector of the total correlation matrix 
$\Lambda$, i.e. $\Lambda\vec{v}\approx n_1\vec{v}$. Any $N\times N$ unitary matrix that 
has $\vec{v}$ as its first column represents a change of basis that puts $\vec{v}$ on 
the first site. The associated transformation of $\Lambda$ will make 
$\Lambda_{11} = n_1$, and zero out the rest of row 1 and column 1.
The matrix of eigenvectors of $\Lambda_{\mathcal{B}}$ would produce such a matrix 
(expanding it to $N\times N$ by putting ones on the diagonal), but this $B\times B$
matrix does not translate well to many-particle gates to use in constructing an MPS.

\begin{figure}
\includegraphics[scale=0.9]{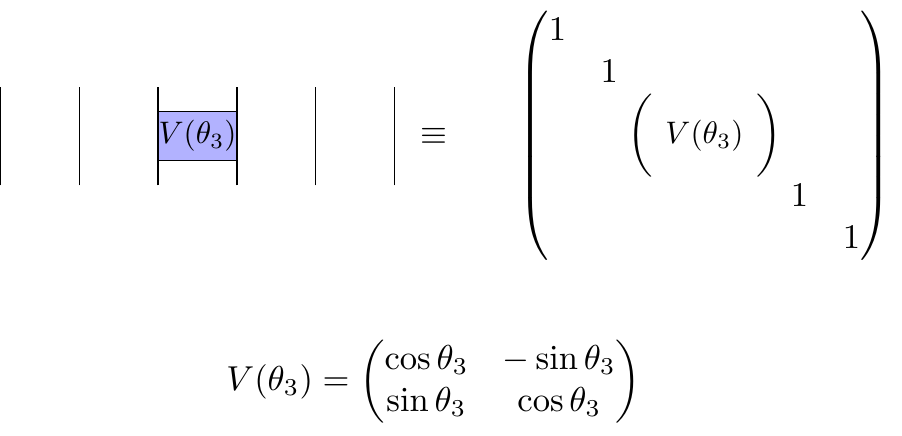}
\caption[]{Definition of a gate used throughout the paper. Example for $N=8$ sites for 
    a gate at site $i=4$. Unless specified otherwise, circuits are in a direct sum space.
    We take the convention that multiplying a matrix from the top by a vector 
    corresponds to multiplying the matrix on the right by a column vector.
    }
\label{gate}
\end{figure}

We now introduce gate/circuit diagrams which apply equally well to simple matrix manipulations
of $\Lambda$ {\it and}  to many-particle tensor networks.  The basic ingredient of the 
diagrams are two site nearest neighbor unitary gates.  
In Figure \ref{gate} we show the relation between a gate and a matrix. 
In the matrix interpretation, an ``MPS" is just a vector with a dimension equal to the
number of sites.
In a later section we show how a gate is interpreted in the many-particle context of
a tensor network.
We consider nearest neighbor gates because these translate to fast MPS 
algorithms---typically, a non-nearest neighbor gate is 
implemented as a set of swap gates to bring the sites together, a nearest neighbor gate, 
followed by swaps to return to the original ordering of the sites, which is much slower 
than a single nearest neighbor gate.  
In the special case that the intermediate sites are in product states,
i.e. bond dimension 1, non-local gates are also inexpensive, and we use these in our
MERA algorithm.

Returning to the task of moving the least entangled state $\vec{v}$ to the first site,
a set of $B-1$ two-site gates suffices.  
The first gate acts on sites $(B-1,B)$, and we label it $V_{B-1}$.  
In general, we take
\begin{eqnarray}
V_i = V(\theta_i) = 
\begin{pmatrix}
\cos\theta_i & -\sin\theta_i \\
\sin\theta_i & \cos\theta_i \\
\end{pmatrix}.
\end{eqnarray}
We choose $\theta_{B-1} = \tan^{-1}(v_B/v_{B-1})$, where $v_i$ is the $i^{\text{th}}$ 
component of the (un)occupied eigenvector of interest $\vec{v}$. With this choice, 
$V_{B-1}$ acting on $\vec{v}^T=\begin{pmatrix}v_1&\ldots&v_{B-1}&v_B\end{pmatrix}$ sets 
the last component, $v_B$, to zero, and produces a new value of $v_{B-1} \to v'_{B-1}$.
In other words, we solve for $\theta_{B-1}$ so that $\vec{v}^TV_{B-1}
=\begin{pmatrix}v_1&\ldots&v_{B-1}&v_B\end{pmatrix}V_{B-1}
=\begin{pmatrix}v_1&\ldots&v'_{B-1}&0\end{pmatrix}$.
Next we rotate sites $(B-2,B-1)$, with $\theta_{B-2} = \tan^{-1}(v'_{B-1}/v_{B-2})$, 
and continue in this fashion.  
The action of all these gates on $\vec{v}^T$ gives $\delta_{i,1}$, so they act to 
change the basis into the one containing $\vec{v}$.

\begin{figure*}
    \subfloat[]{
      \includegraphics[width=0.48\textwidth]{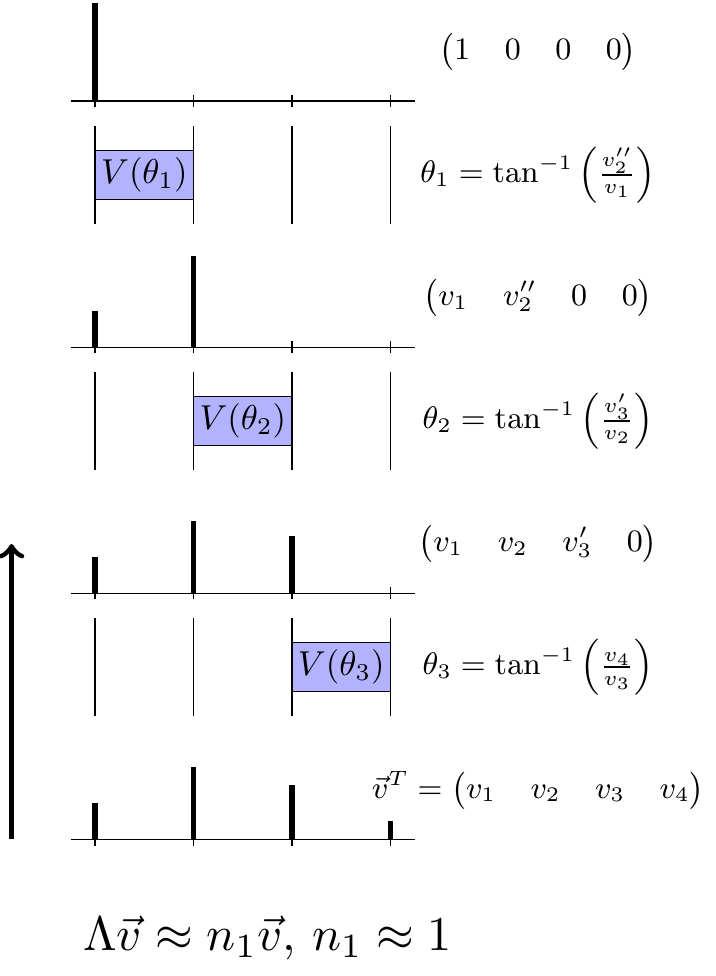}%
      } \ \ \
    \subfloat[]{
      \includegraphics[width=0.48\textwidth]{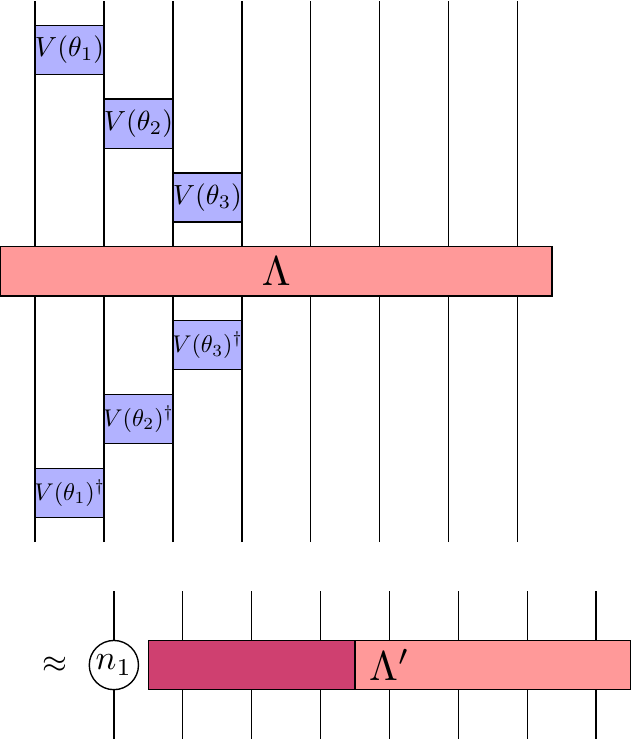}%
      }
    \caption{In Fig.~\ref{mode}(a) we show schematically the 
        procedure to obtain, given an approximate eigenvector $\vec{v}$ of the 
        correlation matrix $\Lambda$, the set of local rotation gates that make up our 
        compressed correlation matrix. The example shown is for a block size $B=4$ and 
        system size $N=8$. Fig.~\ref{mode}(b) shows that, by conjugating the 
        correlation matrix by the gates obtained, the correlation matrix is approximately
        partially diagonalized.}
    \label{mode}
\end{figure*}

We take $V_{\mathcal{B}} = V(\theta_{B-1})V(\theta_{B-2})\ldots V(\theta_1)$.
This procedure is shown schematically for a simple case in Fig.~\ref{mode}(a).
We then apply the gates to $\Lambda$. The transformed correlation matrix 
$V_{\mathcal{B}}^{\dagger}\Lambda V_{\mathcal{B}}$ will have $n_1\approx 1$ or $0$ as 
the top left entry (and nearly 0 in the rest of the entries in the first row and 
first column). A schematic of this transformation is shown in Fig.~\ref{mode}(b). 
We will call the first block $\mathcal{B}_1\equiv\mathcal{B}$. We repeat this procedure 
for $\mathcal{B}_2$, sites $2,\ldots,B+1$, now simply ignoring the first site. For 
half-filled systems, the modes found are likely to alternate between occupied and 
unoccupied because occupied and unoccupied modes will generally be found in pairs when
diagonalizing a block of the correlation matrix. Of course, $B$ does not have to stay the 
same from one block to the next, and in general it is better to set it dynamically to make 
$n_k$ sufficiently close to 1 or 0.  For the last blocks, $B$ is decreased to the remaining
number of sites. After $N$ blocks, we will have approximately diagonalized $\Lambda$. 

\begin{figure*}
    \subfloat[]{
      \includegraphics[width=0.45\textwidth]{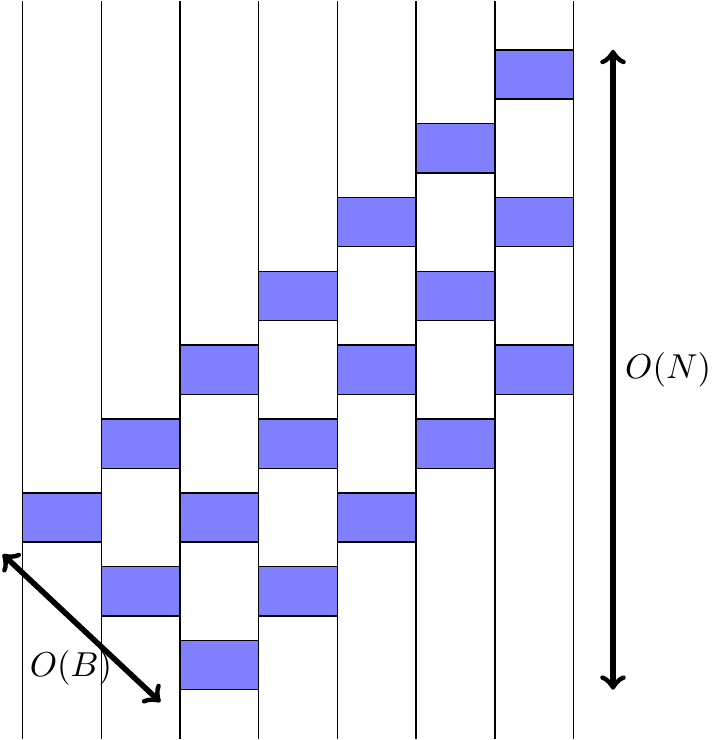}%
      } \ \ \ \ 
    \subfloat[]{
      \includegraphics[width=0.5\textwidth]{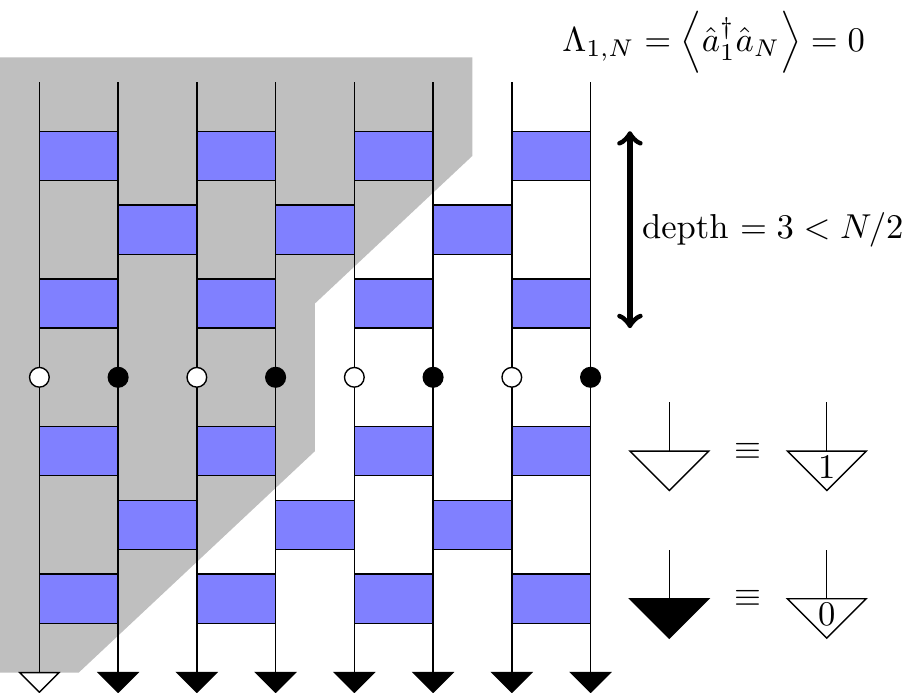}%
      }
    \caption{Fig.~\ref{gate_structure}(a) shows the overall gate structure obtained by 
        the diagonalization procedure. These gates form the total $N\times N$ unitary $V$
        which approximately diagonalizes our correlation matrix $\Lambda$. By conjugating
        a diagonal matrix with the appropriate occupations of 0 or 1 found in the
        diagonalization procedure by this set of gets, we get an approximation for
        the correlation matrix.
        Fig.~\ref{gate_structure}(b) shows an example of the
        correlations allowed by representing the correlation matrix $\Lambda$ with
        a diagonal matrix conjugated by a finite depth circuit of depth $<N/2$. The grey
        area (the ``light cone") represents sites where there can be nonzero correlations
        with the first site. For this circuit depth, there can't be correlations with the
        last two sites. A circuit of depth $\geq N/2$ is required to allow for the
        possibility of nontrivial correlations across the entire system.}
    \label{gate_structure}
\end{figure*}

The overall unitary transformation is $V = V_{\mathcal{B}_1}V_{\mathcal{B}_2}\ldots 
V_{\mathcal{B}_{N-1}}$. The matrix $V$ decomposed into the $2\times 2$ rotation gates 
$\{V(\theta_i)\}$ for $N=8$ and $B=4$ is shown in Fig.~\ref{gate_structure}(a). 
The $N\times N$ unitary approximately rotates our single particle basis from real space 
to what we refer to as the {\it occupation basis}, which is one of the highly 
degenerate eigenbases of the correlation matrix. Conjugating $\Lambda$ by $V$ 
gives us a matrix $V^{\dagger}\Lambda V$ that is nearly diagonal, with $N_F$ values on 
the diagonal close to 1 corresponding to occupied modes and $N-N_F$ values on the 
diagonal close to 0 corresponding to unoccupied modes. In total, the procedure as 
described would require $O(BN)$ nearest neighbor rotations, where $B$ is the largest 
block size needed for the desired accuracy of the representation of the correlation 
matrix.

Writing the $2\times 2$ rotations as gates is very convenient for understanding the
matrix transformations, but more importantly it makes it easy to connect to many-body 
gates and to quantum circuits in general. As a quantum circuit, these gates have a slightly 
peculiar structure. Because of how the diagonalizations overlapped, the circuit has a 
depth of $O(N)$. However, a vertical cut through the circuit only passes through $O(B)$ 
gates. This construction and gate structure is in a certain sense optimal if we limit 
ourselves strictly to circuits with local gates. If we want to represent a correlation 
matrix in a compact way with nearest neighbor gates, we would like to be able to represent 
arbitrary correlations in the system (correlations at all lengths), and in particular,
correlations between the first and last site. In Fig.~\ref{gate_structure}(b), we show a 
circuit which cannot connect the first and last sites because its depth is less than 
$N/2$. Although our gate structure, shown in Fig.~\ref{gate_structure}(a),
has a depth $\sim N$, in fact we can adjust our diagonalization procedure 
slightly to obtain a depth of $\sim N/2$ so our circuit can capture correlations of all 
lengths. This is done by beginning the diagonalization procedure from both the left and 
right side of the system until the blocks meet in the middle. This freedom in where to 
start the diagonalization is similar to the choice of gauge of an MPS. Choosing one 
gauge over another can be useful if we have already performed this procedure for a 
correlation matrix and want to perform it again for another correlation matrix which is 
only locally different from the first one. If we choose the gauge center where the 
correlation matrix has changed, we only need to change a local set of gates.

A generic local circuit of depth $O(N)$ contains $O(N^2)$ gates, and can represent an
arbitrary $N\times N$ single-particle unitary change of basis. 
The low entanglement of physical ground states allows us to represent an $N\times N$ matrix 
with $O(BN)$ one-parameter gates, with $B\ll N$. 
For a gapless system, we know from conformal field theory that the entanglement of a 
subblock $\mathcal{B}$ of $B$ sites varies as $S_B\sim\log(B)$.  
This means that we should be able to capture the entanglement of a critical system of 
$N$ sites with a block size $B\sim\log(N)$.  
If we find that $B\sim\log(N)$, this means that our construction is roughly optimal. 
Fig.~\ref{block_scaling} in Section~\ref{cor_to_gmps_results} shows numerical evidence 
that this is indeed the case.

\subsection{Compressing a Correlation Matrix as a GMERA}
\label{cor_to_gmera}

A MERA tensor network\cite{vidal2007} can represent a 1D critical system using a
constant bond dimension, unlike an MPS.  In our MPS construction, this is
reflected in that $B\sim\log(N)$.
However, we can adjust the diagonalization procedure slightly to obtain a MERA-like 
gate structure with a $B$ which does not grow with $N$. 
The MERA for fermionic Gaussian states 
was first studied in~\cite{evenbly2010-1}, but was only used to study infinite 
translationally invariant systems and required a subtle optimization scheme. Here we will 
show a simpler construction only requiring the tools we have explained so far. 

We begin the procedure in the same way as we did for the GMPS, by diagonalizing the 
block corresponding to sites $1,\ldots,B$ of the correlation matrix. Just as before, for 
a large enough block size we find an occupied or unoccupied mode and rotate into the 
basis containing that mode with $B-1$ local $2\times 2$ gates. Next, instead of 
diagonalizing the block starting at site 2, we instead diagonalize the block corresponding 
to sites $3,\ldots,B+2$, again finding an occupied or unoccupied mode and rotating into 
that basis. The state at site 2 is ``left behind"---it is not a low entangled state,
so we cannot ignore it, but we leave it for a later stage of the algorithm. 
We continue in this manner, diagonalizing blocks starting at odd sites of size 
$B$ to obtain $\sim BN/2$ nearest neighbor gates. Approximately half of the modes are 
fully occupied or unoccupied and are projected out (meaning the associated 
rows and columns in the correlation matrix are ignored in later stages). 
The other half were left behind, and continue as the sites of the next layer of the
gate structure.  By only trying to get $N/2$ unentangled modes in the first layer, the
size of $B$ does not need to grow with $N$, as we show below.

\begin{figure}
    \includegraphics[scale=0.85]{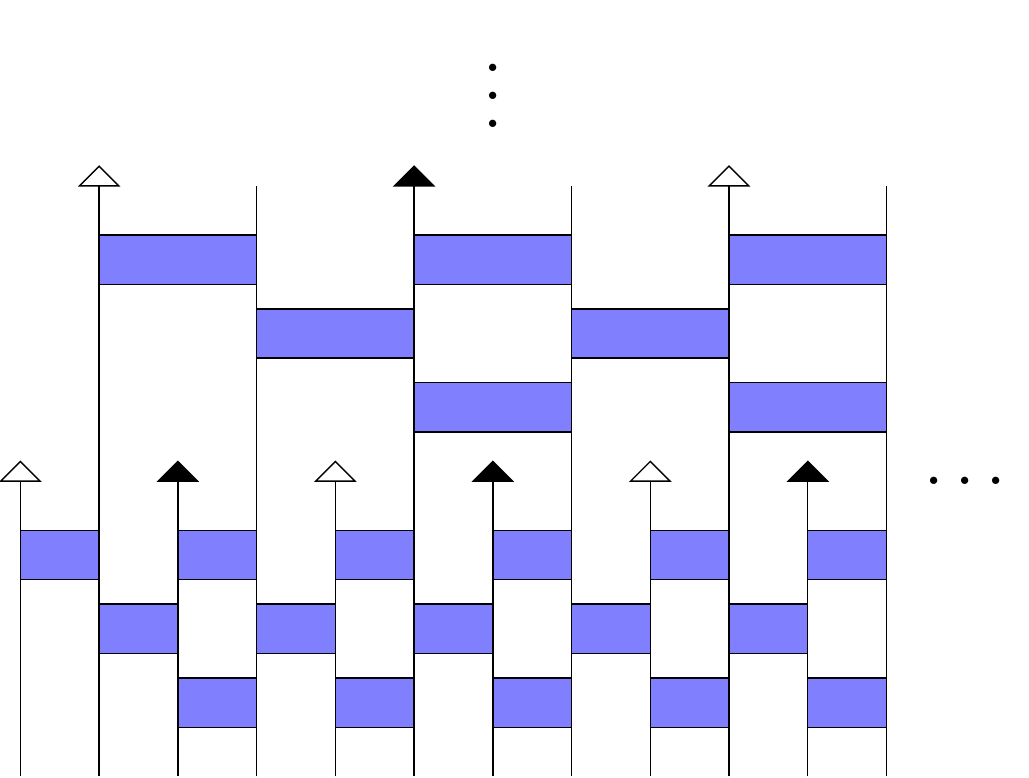}%
    \caption{An example of an alternative diagonalization scheme resulting in a 
        MERA-like gate structure. 
        Here we show a section of the first two renormalization steps, with 12 sites
        shown in the first layer and 6 renormalized sites shown in the second. 
        A block size of $B=4$ is used. For this block size there are two layers of 
        disentanglers and one layer of isometries per level of the MERA.}
    \label{gmera}
\end{figure}

The left-behind sites pass through to the next layer and are interpreted as 
a course-grained version of our original state on only $N/2$ sites. We repeat the same 
procedure for this new course-grained system of $N/2$ sites, starting by diagonalizing the 
subblock of the first $1,\ldots,B$ sites of the new course-grained lattice, finding an 
occupied or unoccupied mode of the course-grained system, and projecting it out. Here,
however, the gate we use to rotate into the basis of the (un)occupied mode are $2\times 2$
nearest neighbor gates in the course-grained lattice, but are actually next-nearest 
neighbor gates acting on the original lattice (if we project out every other site). 
Ordinarily, using next-nearest neighbor gates (or longer range gates at higher levels of 
the MERA) would be costly in the many-body case, requiring swap gates to make them 
effectively nearest neighbor.
However, the projected-out sites are now in product states, meaning that swapping does 
not require significant time.

We repeat the above procedure of projecting out every other effective site and course
graining to a lattice of half the size. All of the sites will be projected out after this 
course-graining is repeated $O(\log_2(N))$ times. Fig.~\ref{gmera} shows the first two 
layers of the resulting gate structure, which looks like a MERA with $B-2$ layers of 
nearest neighbor 2-site disentanglers and a layer of nearest neighbor 2-site isometries. 
The total number of gates in the construction is 
$\sim B(N/2+N/4+\ldots+1)=BN$, the same gate count for a fixed block size $B$ as for the 
GMPS. We call this gate structure, which like our GMPS construction is a compression of an 
$N\times N$ correlation matrix into $\sim BN$ gates, a fermionic Gaussian MERA or GMERA. 
In this figure, triangles denote projections onto the appropriate occupation found. These 
are the modes that are uncorrelated with the rest of the sites and can be ignored in the 
next layer. Some extra gates will be required to project out the leftover sites at the right
end of the system (not shown in Fig.~\ref{gmera}), and there is some flexibility in how to
do this which will change the accuracy of the compression slightly. For example, one could 
use a gate structure similar to the GMPS construction to project out all of the leftover
sites at the end.

How does the block size $B$ of the GMERA compare to that in our GMPS construction? We show 
numerically in Section~\ref{gmera_results} that for a simple gapless Hamiltonian the GMERA 
does indeed produce accurate results with a block size $B=O(1)$, independent of the system 
size, making it much more efficient in the large $N$ limit.

\subsection{Discrete Wavelet Transforms and Fermionic Gaussian MERA}
\label{gmera_wavelet}

We would like to point out the similarity between the MERA gate structure and orthogonal 
wavelet transforms (WT), such as the WTs that produce the well-known Daubechies wavelets
\cite{daubechies1988,daubechies1992}. Of course, the development of wavelets has not
been in a many particle context, and, for now, we restrict ourselves to the
matrix interpretation of the diagrams.  
For compact wavelets, an orthogonal wavelet transform is a local unitary transformation.
It is not usually represented in terms of two-site gates, but this representation turns out
be be particularly convenient.
To be specific, we start with the simplest nontrivial WT, the D4 Daubechies WT. 
This WT is defined by four coefficients $\{a_j\}$ for $j=1,\ldots,4$ which characterize how 
the D4 scaling function is related to itself at different scales through the equation 
$s(x) =\sum_j a_j \sqrt{2}s(2x-j)$.
The matrix form of the WT is given by
\begin{eqnarray}
\begin{pmatrix}
a_1 & a_2 & a_3 & a_4 & 0 & 0 & 0 & \\
a_4 & -a_3 & a_2 & -a_1 & 0 & 0 & 0 &\\
0&0&a_1 & a_2 & a_3 & a_4 &0 &\\
0&0&a_4 & -a_3 & a_2 & -a_1 & 0 &\\
 & & & & & & & &\ddots\\
\end{pmatrix}.
\end{eqnarray}
The $\{a_j\}$ are carefully chosen to ensure orthogonality between scaling functions 
centered at different sites, and to make the scaling functions have desirable 
completeness properties.  
For example, linear combinations of the D4 scaling functions 
centered at different sites, $\{s(x-k)\}$ for integer $k$, fit any linear function, so 
the resulting coefficients are 
$\vec{a}^T=(1+\sqrt{3},3+\sqrt{3},3-\sqrt{3},1-\sqrt{3})/(4\sqrt{2})$.
The orthogonality requirement results in nonlinear equations to solve for the $\{a_j\}$ 
which becomes complicated for higher order. 
The second row of the matrix gives the coefficients that produces wavelets, designed
to represent high momentum degrees of freedom.  In terms of our MERA procedure,
the wavelets are left behind, while the scaling functions propagate to the next level.

\begin{figure}
    \includegraphics[scale=0.75]{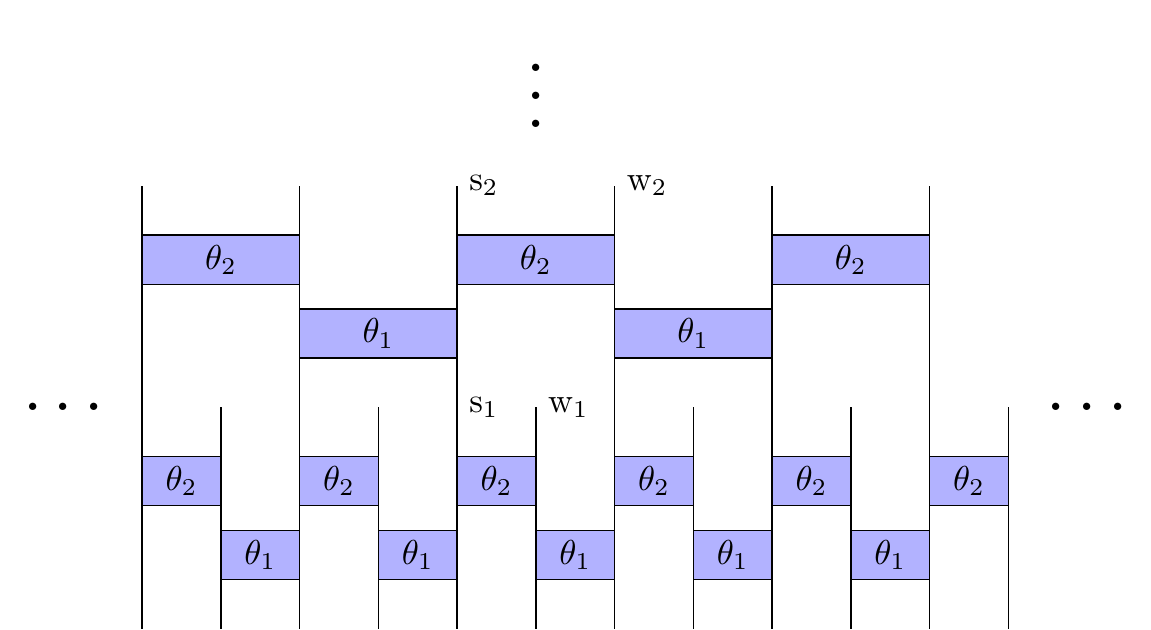}%
    \caption{ Here we show an example of a discrete wavelet transform written in the
        gate notation introduced in this paper. We show the D4 wavelet, which 
        corresponds to a fermionic Gaussian MERA with one layer of disentanglers and one 
        layer of isometries per layer. w$_1$ and s$_1$ (w$_2$ and s$_2$) label wavelet 
        and scaling functions for the first (second) layer. Taking $\theta_1=\pi/6$ and 
        $\theta_2=5\pi/12$, we reproduce the conventional scaling coefficients for
        the D4 WT, $\vec{a}^T=(a_1,a_2,a_3,a_4)=(1+\sqrt{3},3+\sqrt{3},3-\sqrt{3},1-\sqrt{3})
        /(4\sqrt{2})$. }
    \label{wavelet}
\end{figure}

The D4  WT has a very simple gate structure, identical to our MERA structure with $B=3$, 
shown for two layers in Fig.~\ref{wavelet}.
In each horizontal layer of gates, all gates have the same angle.
The D4 WT is specified by only two angles, $\theta_1$ for the bottom layer and $\theta_2$
for the next.
Higher order WTs of this type (e.g. D6, D8, etc.) correspond to larger $B$. 
(For example, the D6 WT looks like Fig. 6).
Given the angles, one gets the $\{a_j\}$ by setting all the top values of the circuit to 
zero except a 1 on one site and applying the $2\times2$ rotations in the layers below. 
The support of the scaling functions is made obvious using the gate structure, as there 
will be $2L$ nonzero values at the bottom of the circuit for $L$ layers of gates.
For the D4 WT, one finds that $\theta_1=\pi/6$ and $\theta_2=5\pi/12$ reproduces the D4 
$\{a_j\}$, up to a trivial reversal of the coefficients.
(A single layer with $\theta_1=\pi/4$ gives the trivial Haar wavelets, which have been 
used previously as a basis for transforming fermionic Gaussian states by Qi~\cite{qi2013}.)
The scaling functions at the larger scales are found by performing the same 
transformation of $L$ layers of gates on the scaling functions found at the previous scale.

\begin{figure}
    \subfloat[Scaling coefficients from gate structure.]{
        \includegraphics[scale=1.0]{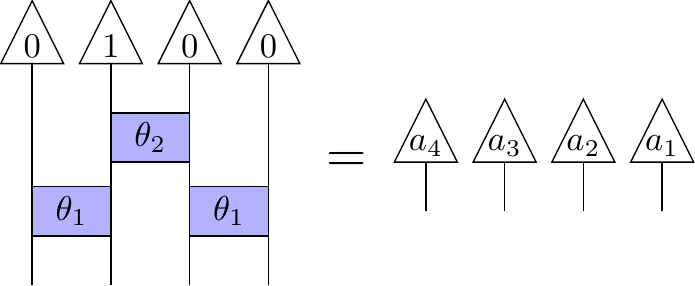}%
        } \\
    \subfloat[Gate structure in (a) written in terms of matrices and vectors.]{
        \includegraphics[scale=0.95]{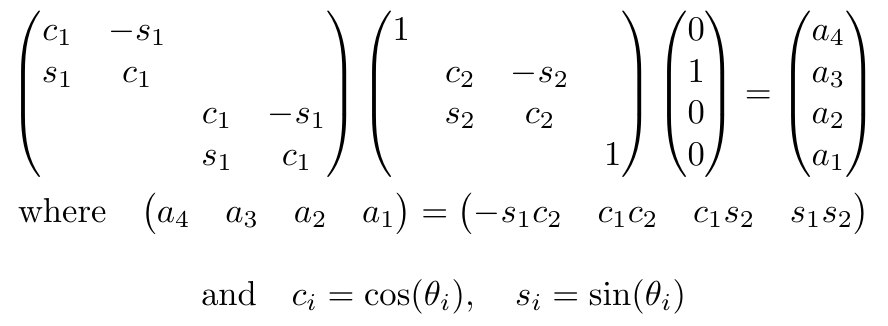}%
        } \\
    \subfloat[Wavelet coefficients from gate structure.]{
        \includegraphics[scale=1.0]{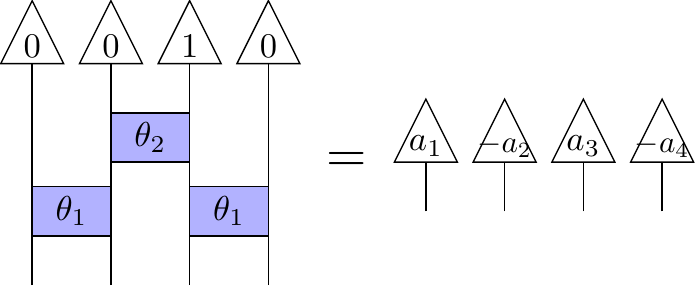}%
        }
    \caption{ Here we show explicitly how to obtain the scaling and wavelet coefficients of
        the D4 WT from the circuit construction. 
        Taking $\theta_1=\pi/6$ and $\theta_2=5\pi/12$, in (a) and (b) we reproduce the 
        conventional scaling coefficients for the D4 WT, $\vec{a}^T=(a_1,a_2,a_3,a_4)
        =(1+\sqrt{3},3+\sqrt{3},3-\sqrt{3},1-\sqrt{3})/(4\sqrt{2})$, up to a trivial
        reversal in the order. 
        In (c) with the same choice of angles we reproduce the conventional wavelet 
        coefficients $(a_4,-a_3,a_2,-a_1)$, again up to a trivial reversal and sign.}
    \label{wavelet_proof}
\end{figure}

In Fig.~\ref{wavelet_proof} we show how scaling coefficients $\{a_j\}$ come from the gate 
structure, applying a vector to the top of the circuit with 1 at the site of a scaling 
function and 0's elsewhere.
In simple wavelet treatments, the wavelet coefficients are obtained from the scaling 
coefficients $\{a_j\}$ as $\{(-1)^{j-1} a_{2L-j+1}\}$ for $j=1,\ldots,2L$. 
Here, they are obtained by shifting the location of the 1 at the top of the circuit, but 
we can show in general that this gives the same result.
This is done by noting that the shift of the 1 to get the wavelet coefficients looks like
a swap at the top of the circuit. 
We can ``pull through" this swap by conjugating each layer of the WT with a transformation 
that reverses the order of the sites.
This conjugation also negates the angles in the circuit. 
It leaves a site reversal at the bottom of the circuit, reversing the order of the 
coefficients. 
The angle negation negates the sine terms, leading to the same coefficients except with 
every other one negated, since every other site will have an even or odd number of 
$\sin(\theta_i)$ multiplied together. 

Given an arbitrary set of $\{a_j\}$, we can use the same procedure that brought $\vec v$
to the first site in our GMPS procedure to find all the angles defining the WT, i.e 
$\vec v = \vec a$.  
Thus, any compact orthogonal WT of this general type can be represent by a simple gate 
structure.
Because wavelets are much easier to understand than generic many particle wavefunctions, 
the connection between MERA and wavelets may help provide intuition that helps one 
understand MERA.

\subsection{Forming the Many-Body MPS from the GMPS (or GMERA)}
\label{gmps_to_mps}

\begin{figure*}
\includegraphics[scale=1.25]{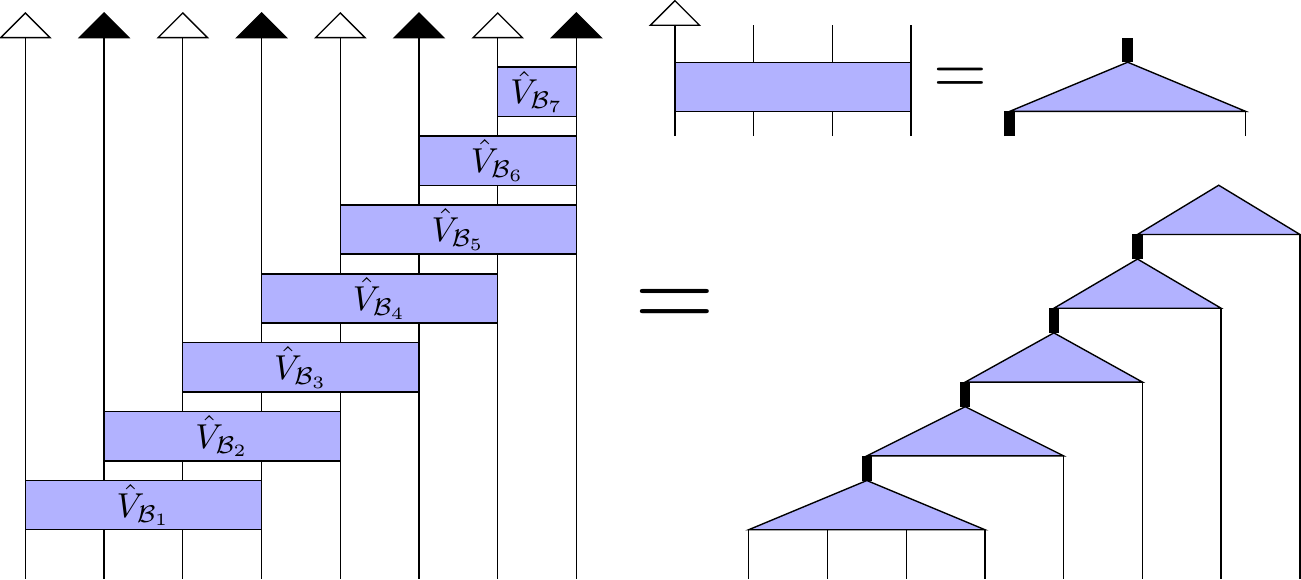}
\caption[]{Tensor diagram showing the structure of gates $\{\hat{V}_{\mathcal{B}_i}\}$ 
    for $i=1,\ldots,N-1$ obtained in our procedure and how they contract to form an MPS. 
    Here we show a system with $N=8$ sites and a block size $B=4$. The diagram on the 
    right shows that once the sites are rotated into a basis where one of the modes is 
    occupied or unoccupied (generally with some alternating pattern), the fully occupied
    or unoccupied modes can be projected out. The transformations 
    $\{\hat{V}_{\mathcal{B}_i}\}$, including the projections, can be directly 
    interpreted as the tensors composing the MPS representation of our many-body state 
    if we do an exact contraction, or we can apply them as a set of gates as explained 
    in the text.}
\label{mps}
\end{figure*}

For a number conserving real Hamiltonian $H$, the many particle 
unitary gate $\hat{V}_i$ corresponding to the single particle rotation $V_i$, 
in the basis 
$\{\ket{\Omega},
\hat{a}^\dagger_i\ket{\Omega},\hat{a}^\dagger_{i+1}\ket{\Omega},
\hat{a}^\dagger_i\hat{a}^\dagger_{i+1}\ket{\Omega}\}$, is
\begin{eqnarray}
[\hat{V}_i] = [\hat{V}(\theta_i)] =
\begin{pmatrix}
1 & 0 & 0 & 0 \\
0 & \cos\theta_i & \sin\theta_i & 0 \\
0 & -\sin\theta_i & \cos\theta_i & 0 \\
0 & 0 & 0 & 1 \\
\end{pmatrix}.
\label{mbgate}
\end{eqnarray}
This reinterpretation of the gates is the only change need to make our matrix
gate structures act on the many particle Hilbert space.

Say we have compressed the correlation matrix of a pure fermionic Gaussian state as a GMPS. 
To create the MPS representation of this state, we begin with a product state, with each 
site being occupied or unoccupied, with the occupations given by $n_k$ obtained in our 
diagonalization procedure (set to 1 or 0 for $n_k\approx 1$ or 0). We then apply, one 
by one, all of the nearest neighbor gates $\{\hat{V}_i\}$ (the many-body gates
corresponding to the gates $\{V_i\}$ obtained with Eq.~\ref{mbgate}) in the opposite order 
in which they were obtained with our diagonalization procedure. 
The repeated application of gates is similar to the time-evolving block decimation 
(TEBD) algorithm\cite{vidal2004} or the time dependent DMRG algorithm\cite{white2004}, 
but the pattern of gates and ordering is different.  We apply the two body gates by 
moving the center of the MPS to the location of the gate, contract the gate with the 
two relevant tensors in the MPS, and then form the new MPS by performing a singular
value decomposition (SVD), with possible truncation of states by throwing out states 
with small singular values. 

We can also form the MPS from our GMERA construction in a similar manner. 
However, instead of starting with a full product state, we start with the gates at the top of 
the MERA and work our way down, including only the sites that have been touched by a gate
at that level or above.
When a site is added, it starts as a completely occupied or unoccupied state,
and is immediately mixed with another site by a gate.
The number of sites involved roughly doubles with each layer, and after $O(\log_2(N))$
layers of the MERA we have our MPS approximation for the entire system.
Again, we can truncate as needed by throwing out low weight states after
the SVD as we work our way down.

Returning to the MPS construction, the tensors of the MPS could also be constructed 
directly by contractions of the gates as shown in Fig.~\ref{mps}. 
In this diagram the small black and white triangles signify projectors onto the appropriate 
occupations found, while the thick lines signify combined internal indices which form the 
internal bonds of the MPS. 
From this perspective it is easy to see that picking a block size $B$ for diagonalizing the 
correlation matrix would correspond to an MPS with a bond dimension of $\chi = 2^{B-1}$. 
We find it simpler and more efficient to apply the gates layer by layer instead of this
method.  Layer by layer, it is natural to truncate the MPS with SVDs during the construction,
and this can lead to an MPS with a smaller bond dimension than $2^{B-1}$ for the required
accuracy.  The SVD truncation takes one out of the manifold of Gaussian states, where the
greater freedom for a fixed bond dimension allows one to find a state which is
closer to the desired Gaussian state than one could within the Gaussian manifold.
However, one should pick a block size so that $2^{B-1}$ is as close to the target
bond dimension as possible.

We can adapt our circuits to complex quadratic Hamiltonians, 
where the gates are of the same form but the $2\times 2$ submatrix rotating the
singly occupied subspace is a general matrix in SU(2) parameterized by two angles. Even
more generally, we can extend this procedure to quadratic Hamiltonians with pairing
terms to compress BCS states, where the gates required are not just number conserving but 
general parity conserving gates (so they involve mixing of unoccupied and doubly occupied 
subspaces of the 2 sites of interest). This matrix would in general be parameterized by 
5 angles (one matrix in SU(2) rotating the singly occupied subspace, one matrix in SU(2)
rotating the empty and doubly occupied subspaces, and a relative phase). This form 
of gates has been studied previously in the context of classically simulating quantum 
circuits using the matchgate formalism; see for example\cite{valiant2001,jozsa2008}.

\section{Numerical Results}
\label{results}

Here we show numerical results for the algorithms we presented.
In order to study systems that are both gapless and gapped, we study a simple model, the 
Su-Schrieffer-Heeger model~\cite{su1979}. 
This is a model of 1D spinless fermions hopping on a lattice with staggered hopping 
amplitudes, $t_1$ and $t_2$. The Hamiltonian is
\begin{eqnarray}
\hat{H}_{SSH}=\sum_{i=1}^{\frac{N-1}{2}} (t_1 \hat{a}_{2i-1}^{\dagger}\hat{a}_{2i}
    +t_2 \hat{a}_{2i}^{\dagger}\hat{a}_{2i+1}+h.c.).
\end{eqnarray}
We will take $t_1=-t\left(1+\frac{\delta}{2}\right)$ and 
$t_2=-t\left(1-\frac{\delta}{2}\right)$.
The model has an energy gap in the bulk between the ground state and first excited state 
of $\Delta=2|\delta|t$ in the thermodynamic limit ($N \rightarrow \infty$). 
With open boundary conditions, it can contain exponentially decaying zero energy modes 
localized on the ends of the chain. 

\subsection{Results for Compressing a Correlation Matrix as a GMPS}
\label{cor_to_gmps_results}

\begin{figure}
\includegraphics[scale=0.715]{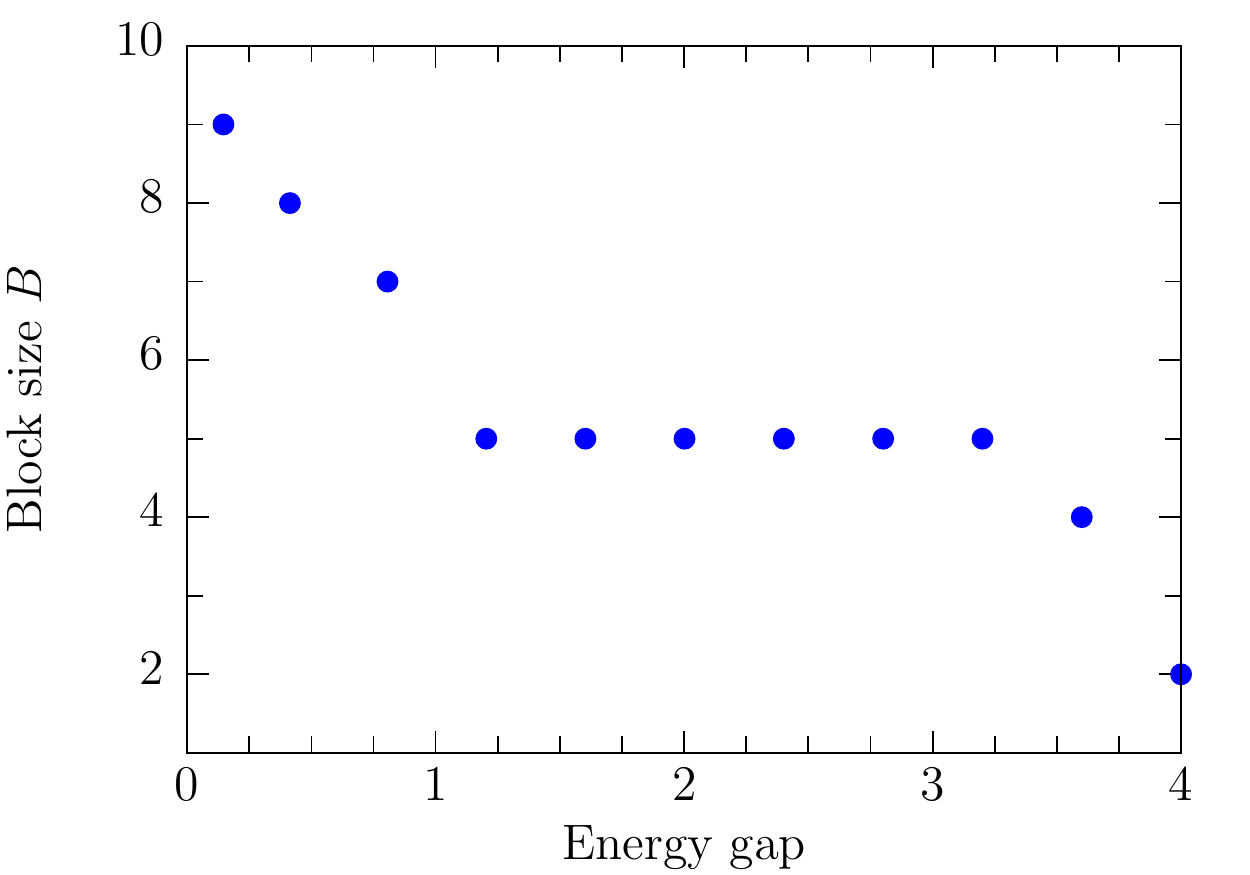}
\caption[]{Block size required to obtain a relative error in the total
    energy of less than $10^{-6}$ as a function of the calculated energy gap (in
    units of $t$) for $N=128$ sites.}
\label{block}
\end{figure}

\begin{figure}
\includegraphics[scale=0.715]{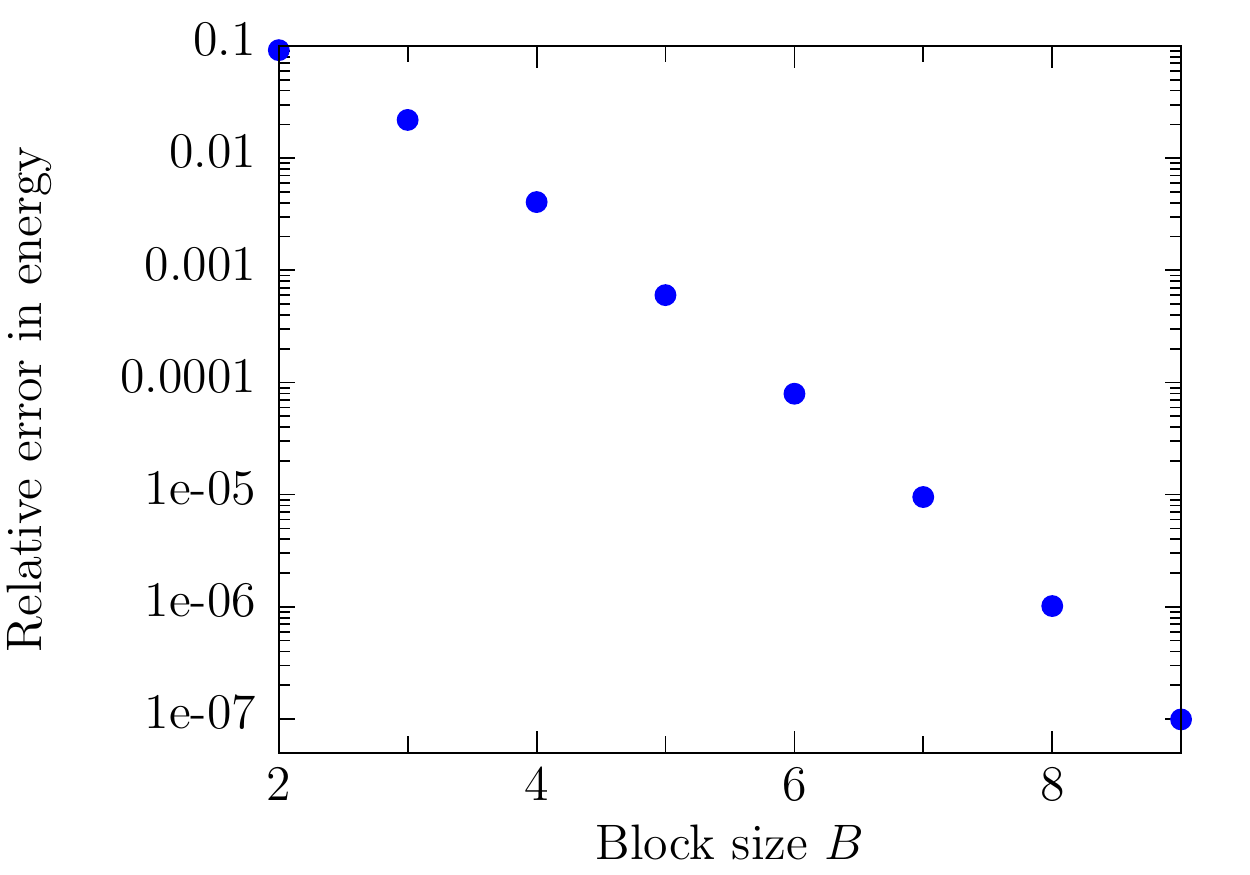}
\caption[]{Relative error in the total energy as a function of the block
    size $B$ for $N=128$ sites and $\delta=0$.}
\label{error}
\end{figure}

We start with a simple test of obtaining the GMPS compression of the ground state 
correlation matrix of the SSH model for $N=128$ lattice sites for various energy gaps at 
half filling ($N_F=N/2$). 
We analyze the range of $\delta$ from 0 to $2$. 
The ground state for $\delta=0$ is (approximately) gapless while for $\delta = 2$ it is fully 
gapped (the chain uncouples).  
Fig.~\ref{block} shows the block size required to obtain a GMPS with a relative 
error in the total energy of less than $10^{-6}$ as a function of the calculated 
energy gap. The exact ground state energy and energy gap are calculated by 
diagonalizing the hopping Hamiltonian $H_{SSH}$. This corresponds to the accuracy of
the MPS representation of the ground state if the GMPS written with many-body gates is
contracted with no further truncation of the MPS, so a GMPS block size $B$ 
corresponds to an MPS of bond dimension $\chi = 2^{B-1}$ (which is why the 
block size remains constant for intermediate energy gaps). The plot shows, as 
expected, that the block size required decreases as the energy gap is increased.
Fig.~\ref{error} shows, for the case $\delta=0$ (where the energy gap, due to the
finite size, is 0.146088$t$), the relative error in the energy as a function of the block 
size.

\begin{figure}
    \subfloat[$\delta=0.4$ (energy gap $\approx 0.806135 t$)]{
      \includegraphics[width=0.475\textwidth]{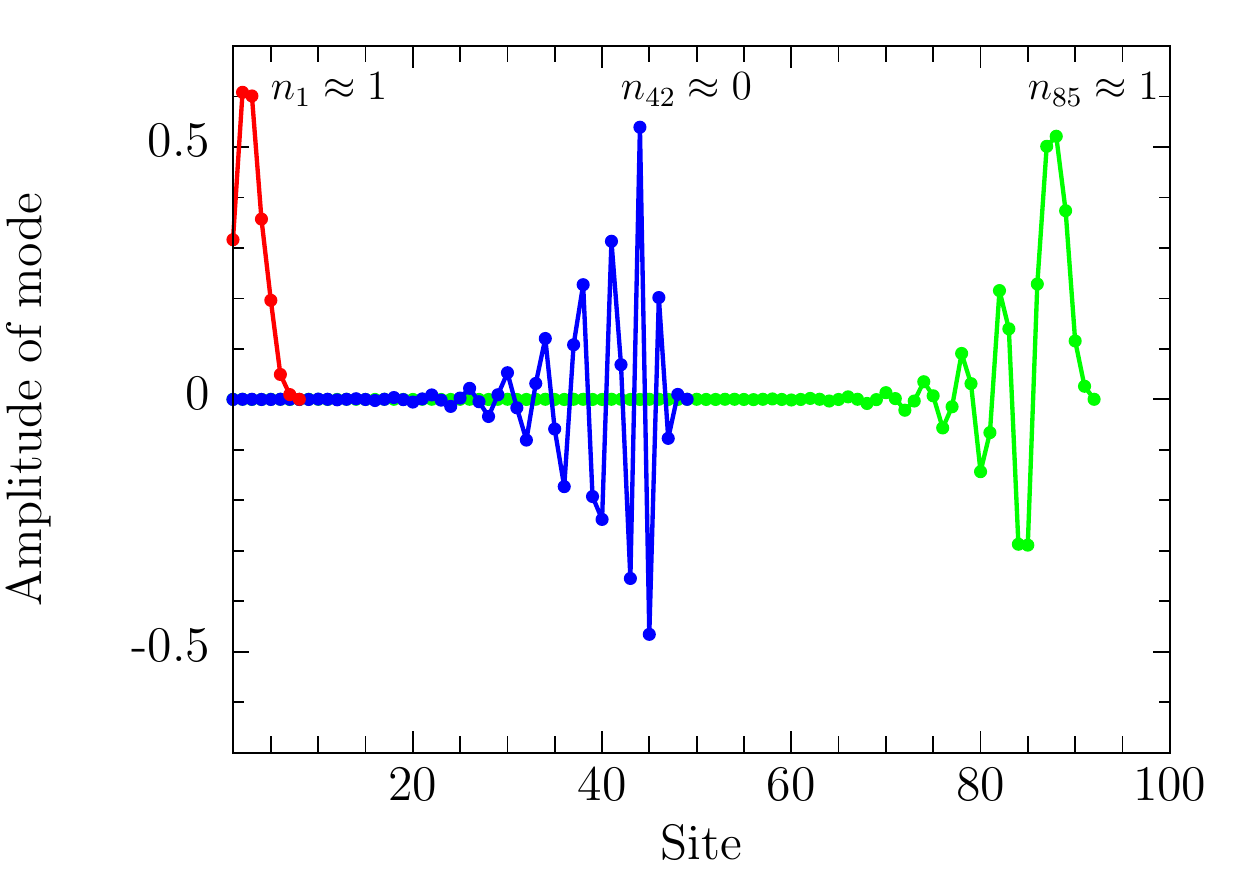}%
      } \\
    \subfloat[$\delta=0$ (energy gap $\approx 0.146088 t$)]{
      \includegraphics[width=0.475\textwidth]{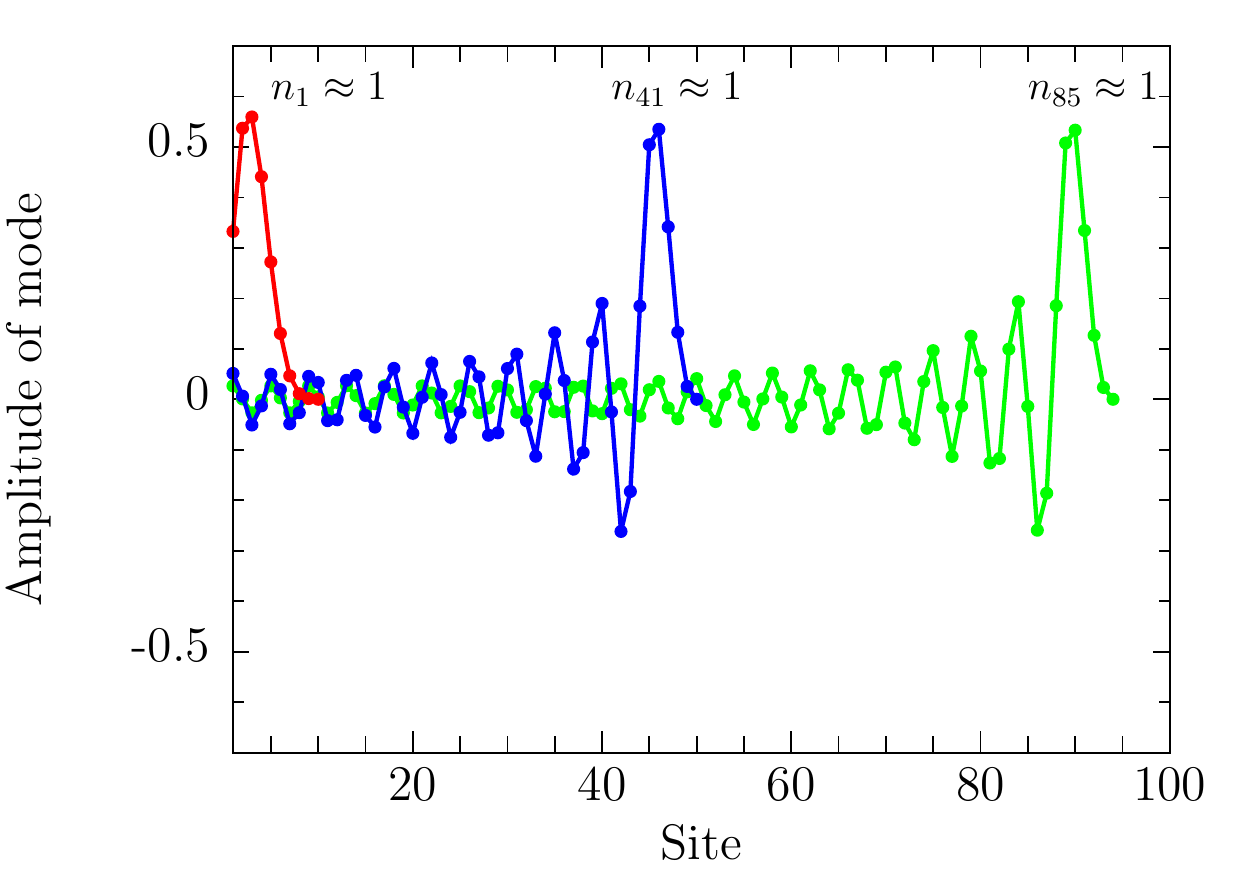}%
      }
    \caption{Examples of occupied and unoccupied modes found in the diagonalization
        process. Fig.~\ref{modes}(a) shows occupied/unoccupied modes for 
        $\delta=0.4$ (energy gap $\approx 0.806135 t$). Fig.~\ref{modes}(b)
        shows occupied/unoccupied modes for $\delta=0$ (energy gap $\approx 
        0.146088$).}
    \label{modes}
\end{figure}

\begin{figure}
    \subfloat[$\delta=0.4$ (energy gap $\approx 0.806135 t$)]{
      \includegraphics[width=0.475\textwidth]{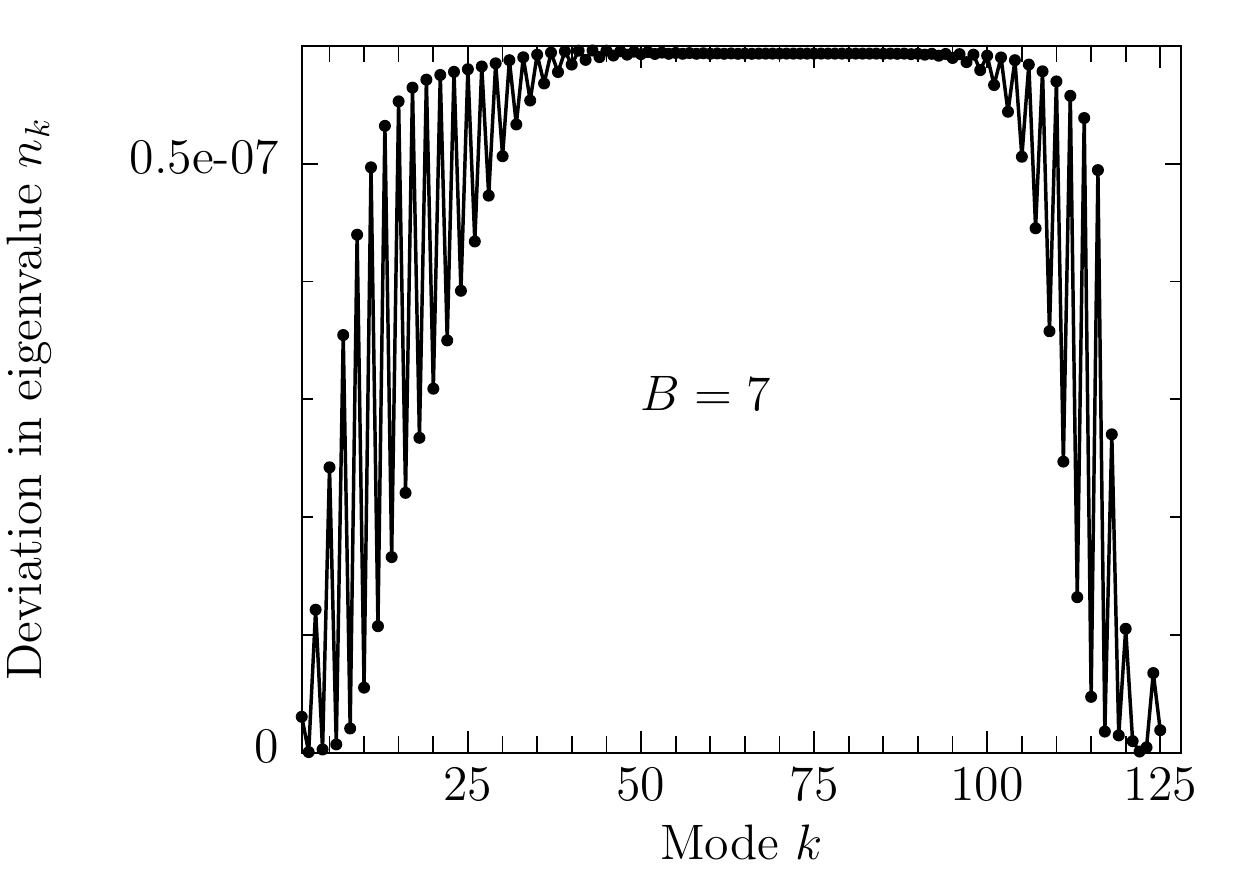}%
      } \\
    \subfloat[$\delta=0$ (energy gap $\approx 0.146088 t$)]{
      \includegraphics[width=0.475\textwidth]{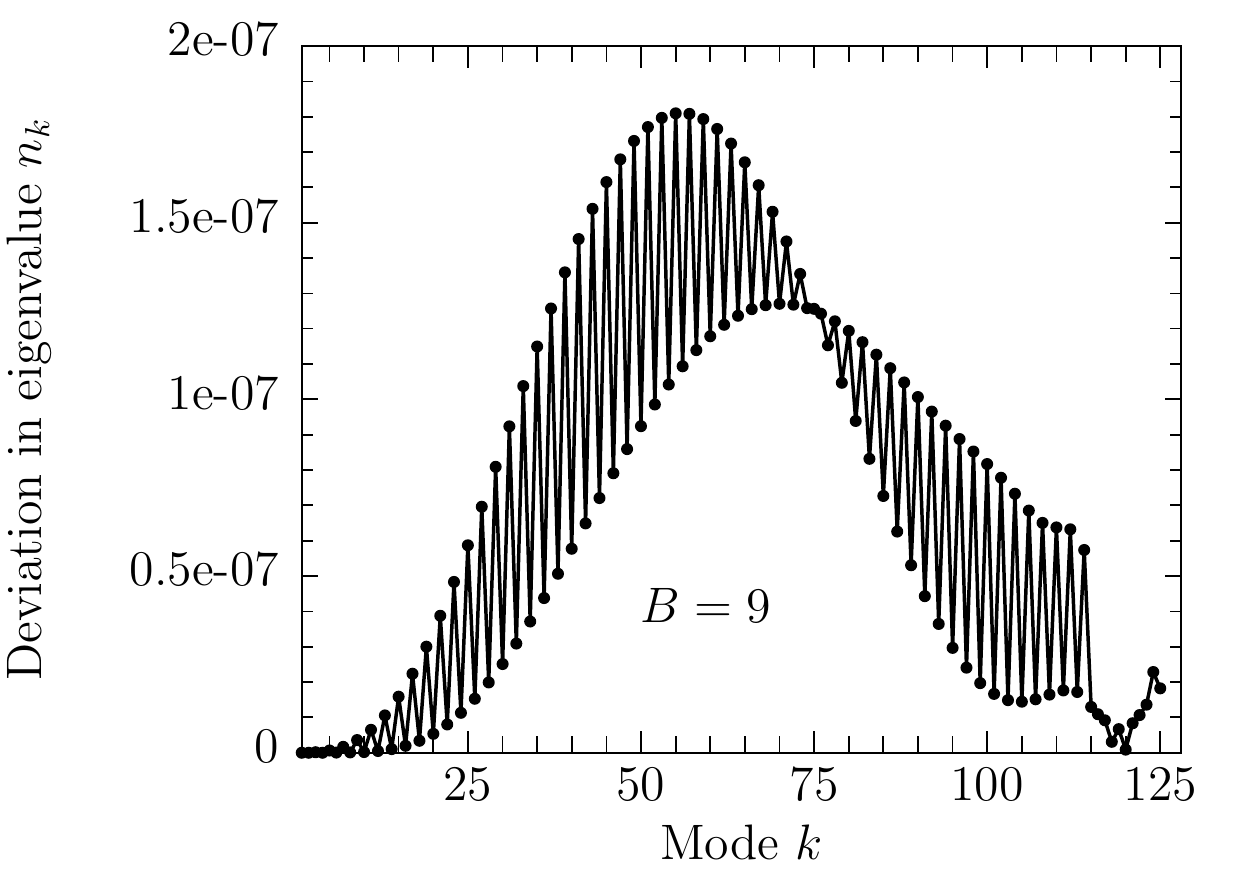}%
      }
    \caption{Examples of deviations in occupations at the end of the diagonalization
        procedure for $N=128$ sites. Fig.~\ref{zeta}(a) shows errors in the 
        occupations for $\delta=0.4$ (energy gap $\approx 0.806135 t$). 
        Fig.~\ref{zeta}(b) shows errors in the occupations for $\delta=0.0$ 
        (energy gap $\approx 0.146088 t$).}
    \label{zeta}
\end{figure}

Fig.~\ref{modes} shows examples of the modes obtained with the procedure, both filled 
and unfilled, for a small gap and a larger gap. The modes are seen to be localized 
for the case of the larger gap, and extend throughout the system for the smaller gap. 
The unfilled modes follow the same decay as the filled modes but oscillate more, 
since they are above the Fermi sea and are therefore higher in energy. Fig.~\ref{zeta} 
shows for the same two gaps the deviation in the eigenvalues $n_k$ from 0 or 1 obtained 
during the diagonalization procedure. For the case of the larger gap, this error 
saturates to its maximum quickly for modes near the middle of the system, while for the 
smaller gap, the error increases more slowly due to the longer range correlations.

\begin{figure}
    \includegraphics[scale=0.7]{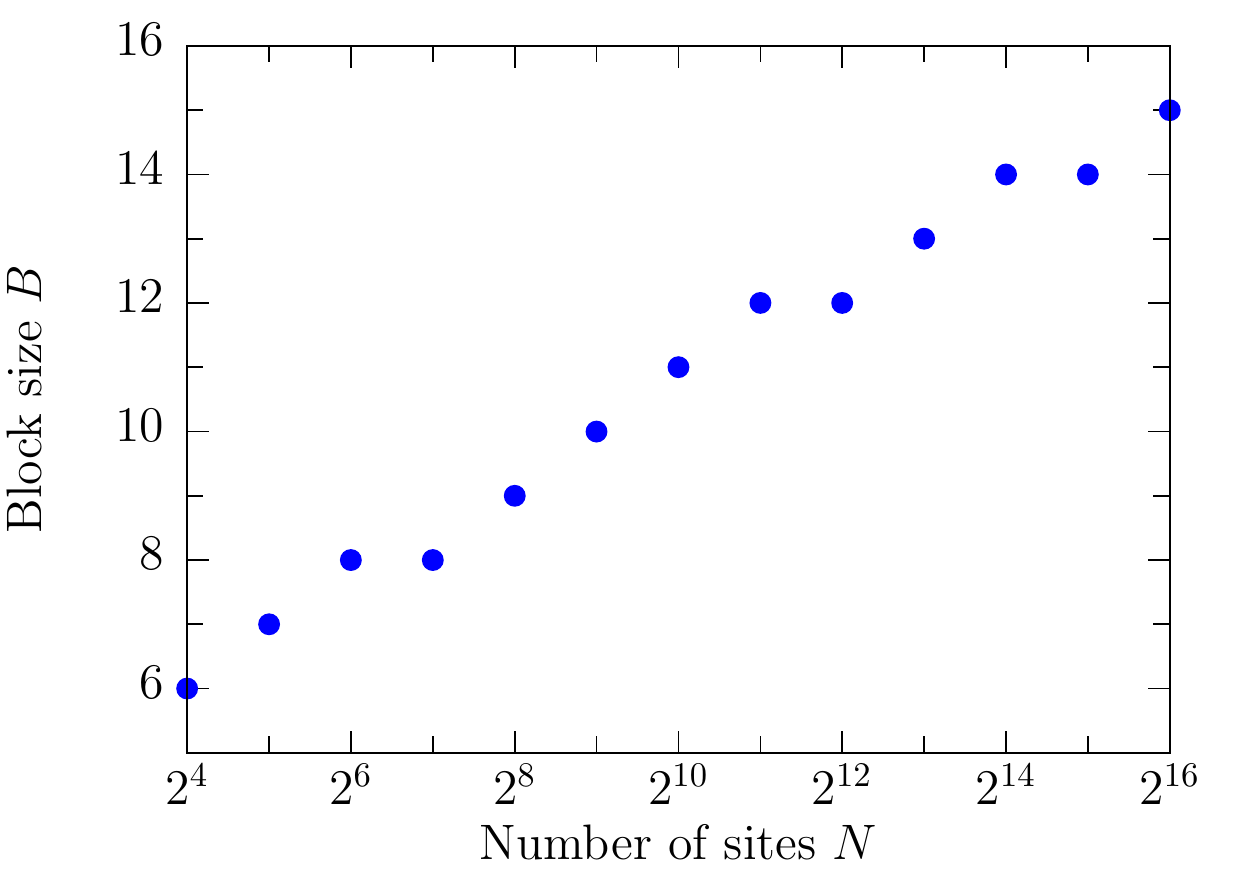}%
    \caption{Block size $B$ needed for a relative error in the energy of $<10^{-6}$
        as a function of number of sites $N$ for spinless, gapless fermions with
        open boundary conditions at half filling. As expected from arguments about the 
        entanglement of a critical system, we find $B\sim\log(N)$, tested up to 
        $N=2^{16}=65536$ sites (note the $\log$ scale on the x axis). 
        To study systems of this size and avoid the $O(N^3)$ 
        diagonalization of the hopping Hamiltonian, we obtain the correlation matrix
        using the GDMRG algorithm as explained in Appendix~\ref{appendix2}. }
    \label{block_scaling}
\end{figure}

In Fig.~\ref{block_scaling} we analyze the block size scaling with system size $N$ for
the gapless case ($\delta=0$). As we expect from arguments about entanglement made at the 
end of Section~\ref{cor_to_gmps}, the scaling is found to be $B\sim\log(N)$. This is 
the expected scaling for a critical 1D system. We can see that with this procedure we 
can analyze very large systems, up to $N=2^{16}=65536$ sites, even for gapless free
fermions. To avoid storing correlation matrices this large, we begin with a very accurate
compressed correlation matrix as a GMPS using the GDMRG algorithm presented in 
Appendix~\ref{appendix2}. With GDMRG, we begin with a state with a relative error in the
energy of $<10^{-10}$. For $N=65536$ this requires a block size of $B=22$. We then obtain
the local correlation matrix for the block we are interested in using the gates from
this accurate compression, and use it to obtain a less accurate compression with a 
smaller block size. This procedure should lead to, for a given block size, a more accurate
overall state than one that would be obtained directly from GDMRG, because GDMRG optimizes
the energy which only depends on very local correlations.

\subsection{GMERA Results}
\label{gmera_results}

\begin{figure}
    \includegraphics[scale=0.7]{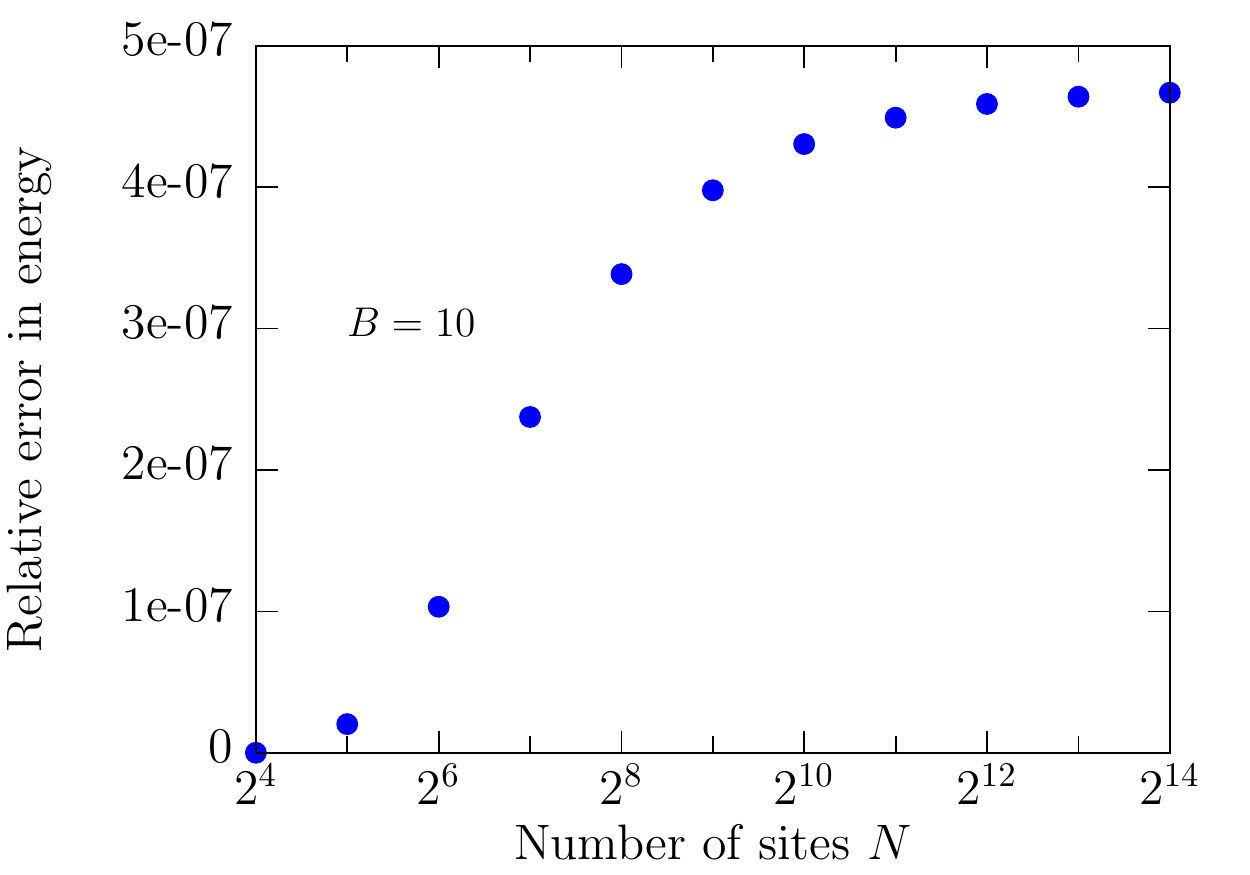}%
    \caption{ Relative error in the energy for the proposed GMERA construction for 
        increasing number of sites for a block size $B=10$. The system analyzed is the 
        ground state of free fermions hopping on a lattice with open boundary conditions.
        All errors are below $10^{-6}$. As expected for a MERA, the error is seen to 
        saturate for large $N$, indicating a fixed block size is sufficient to obtain an 
        accuracy $<10^{-6}$ up to very large system sizes. }
    \label{gmera_error_scaling}
\end{figure}

Here we present results for compressing a correlation matrix as a GMERA using the
procedure presented in Section~\ref{cor_to_gmera}.
We show the relative error in the energy for increasing number of sites for 
$B=10$ in Fig.~\ref{gmera_error_scaling}. 
We see that for this block size, the error stays below $10^{-6}$ for systems up to 
$N=2^{14}=16384$ and in fact appears to saturate at high number of sites (the change in the 
relative error in the energy approaches 0 for larger system sizes). 
This is in stark contrast to the GMPS, where a block size $B\sim\log(N)$ was required to 
obtain a fixed accuracy, as shown in Fig.~\ref{block_scaling}. 
Instead, the GMERA obtains the same accuracy with constant block size $B$ as shown in 
Fig.~\ref{gmera_error_scaling}. 
The GMPS obtains the given accuracy with the same or smaller block size up to $N=512$, after 
which it requires a larger block size than the GMERA to obtain the same level of accuracy. 
As we mentioned earlier, this is made possible partially because the GMERA structure involves 
nonlocal gates. 

\subsection{GMPS to Many-Body MPS Results}
\label{gmps_to_mps_results}

\begin{figure}
    \includegraphics[width=0.5\textwidth]
        {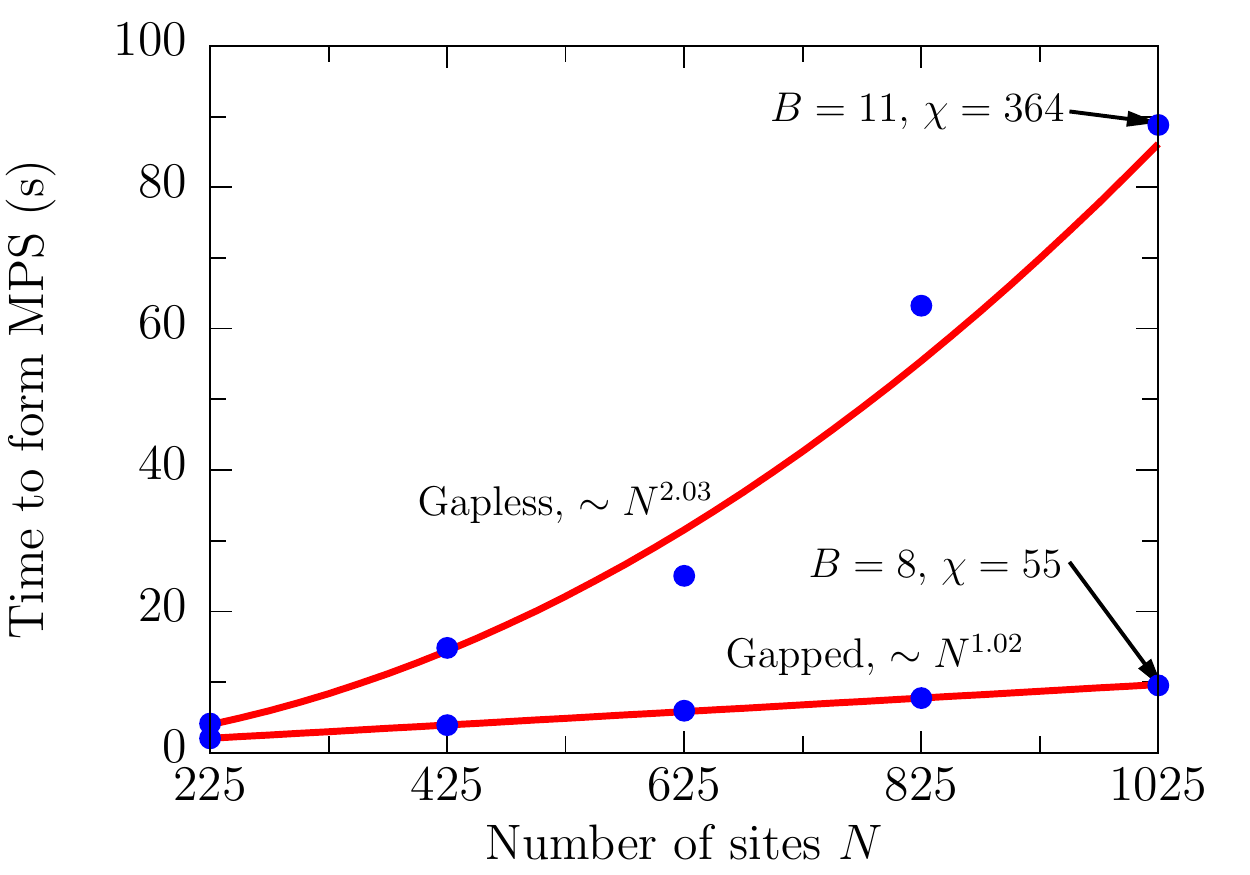}%
    \caption{ A plot of the time to form the MPS approximation of gapped and gapless free 
        fermion ground states at half filling as a function of sites $N$ using gates
        from a GMPS. The bond dimensions are chosen large enough such that the relative 
        errors in the energy of the MPS are below $10^{-6}$. The block size of the GMPS 
        used to form the MPS are the minimum required to obtain a GMPS with a relative
        error in the energy of $10^{-6}$. A cutoff in the singular values of the SVD of 
        $10^{-11}$ was used when applying the gates to form the MPS using the method 
        described in Section~\ref{gmps_to_mps}.
        For the gapped case, the SSH model with $\delta=0.1$ is used, corresponding to an 
        energy gap of $\Delta\approx 0.2t$ (exact as $N\rightarrow \infty$).
        }
    \label{timing}
\end{figure}

Plots of the time it takes to form the MPS of the ground state of a gapless free fermion 
system for up to $N=1024$ sites using the method presented in Section~\ref{gmps_to_mps} 
are shown in Fig.~\ref{timing}. 
As expected, the time it takes for a gapless system is polynomial in the system size $N$,
while it is approximately linear in $N$ for a gapped system. The SSH model is used with 
$\delta=0.1$ or an energy gap $\Delta\approx 0.2 t$.
The time to form the gapless ground state is only a modest polynomial in $N$, $\sim N^{2.03}$,
while as we expect from arguments about entanglement the time to form the gapped ground
state is very nearly linear in $N$, $\sim N^{1.02}$, because the block size and bond dimension
required to obtain the specified accuracy are constant for all $N$ shown ($B=8$ and 
$\chi=55$).
With this method, a gapless ground state of $N=1025$ sites with a relative error in the
energy of $<10^{-6}$ can be formed on a laptop in only $\sim 90$ seconds. 

An interesting point to emphasize is the quality of the compression. The GMPS for the 
gapless ground state on $N=1025$ sites requires a block size of $B=11$ to obtain a relative 
error in the energy of $<10^{-6}$. 
Naively, turning these gates into many-body gates and contracting the network
(forming the MPS directly from the GMPS with no truncation) as explained earlier leads to
a bond dimension of the MPS of $\chi=2^{B-1}=2^{10}=1024$. However, applying the gates as 
described and using a cutoff of the singular values of $10^{-11}$ leads to the formation
of an MPS approximation of the fermionic Gaussian ground state, still with a relative error
in the energy of $<10^{-6}$, with a bond dimension of only $\chi=364$. This is a result
of the fact that our GMPS only explores the manifold of fermionic Gaussian states limited 
to the specified block size. 
On the other hand, the MPS approximation of the Gaussian state we form 
from this GMPS is able to explore the entire manifold of MPS's up to the allowed bond 
dimension (and particle number if symmetric tensors are used, as we do here), so through the 
SVD we are able to compress the state quite efficiently beyond what we initially might expect.

\section{Conclusion}

We have presented an efficient, numerically stable, and controllably accurate way to 
compress a correlation matrix into a set of $2\times 2$ unitary gates. 
From these gates, we have also presented a method for easily and efficiently forming the 
MPS approximation of a fermionic Gaussian state.
We explained the procedure in detail for the ground state of a generic number 
conserving Hamiltonian. We then presented results for the SSH model, a 1D chain of 
fermions with staggered hopping. We showed examples of the accuracy and block 
sizes needed for different gaps of the model. We hope this method can be used as a 
simple, efficient and reliable procedure for directly preparing many states of 
interest, either by creating starting states to aid DMRG calculations or preparing a 
particular ansatz as an MPS. We also presented one example of how the procedure can
be modified to obtain different gate structures, in this case one that is related to
the MERA. However, there are other possibilities to be explored, such as gate structures
more directly suited for systems with 2 spatial dimensions, periodic boundary conditions,
as well as how the method might be applied to study thermal fermionic Gaussian states. 
In addition, we presented how discrete wavelet transforms can be described very simply with 
the gate structure notation we introduced in this paper.

The method is easily generalized to cases beyond the one presented here. As we touched 
upon earlier, it can be generalized to the case of BCS states, the ground states of hopping 
Hamiltonians that include pairing terms.
In this case, the correlation matrix in the Majorana basis can be written in terms of an 
anti-symmetric matrix which can be approximately block diagonalized
by $\sim 5BN$ local $2\times 2$ rotation gates, which are turned into
nearest neighbor parity-conserving many-body gates. The case of spinless fermions 
was presented, but spinful fermions are a simple generalization.

\begin{acknowledgments}
We would like to thank Glen Evenbly for many helpful discussions and comments on the 
manuscript.
We would also like to acknowledge input from Mike Zaletel and Garnet Chan.
This material is based upon work supported by the National Science Foundation Graduate 
Research Fellowship under Grant No. DGE‐1144469. Any opinion, findings, and 
conclusions or recommendations expressed in this material are those of the authors(s) 
and do not necessarily reflect the views of the National Science Foundation. This work 
was also supported by the Simons Foundation through the many electron collaboration.
\end{acknowledgments}

\appendix
\section{Calculation of the Entanglement Entropy of a Fermionic Gaussian State}
\label{appendix1}

In this section we give a simple, self-contained derivation for Eq.~\ref{entropy},
the entanglement entropy for a block of a free fermion system. 
Assume the block of interest $\mathcal{B}$ is the first $B$ sites. 

Gaussian states have expectation values that obey Wick's theorem. 
This means that the expectation value of any operator contained within the block is 
specified if we know subblock $\mathcal{B}$ of the correlation matrix, 
$\Lambda_{\mathcal{B}}$. 
This implies that the many-body density matrix of the block $\hat{\rho}_{\mathcal{B}}$ is 
also uniquely specified by $\Lambda_{\mathcal{B}}$. 
The entanglement entropy on block $\mathcal{B}$, defined as 
$S_B[\hat{\rho}_{\mathcal{B}}]=-\Tr[\hat{\rho}_{\mathcal{B}}\log(\hat{\rho}_{\mathcal{B}})]$,
does not change under general unitary transformations within the block. 
Thus, we can perform the single particle unitary transformation of basis that makes 
$\Lambda_{\mathcal{B}}$ diagonal, with diagonal
entries $n_b = \left<\hat{a}^{\dagger}_b\hat{a}_b\right>$ for $b\in 1,\ldots,B$. The
$n_b$ uniquely specify the reduced density matrix of the block, so the entanglement
is a universal function of $\{n_b\}$:
\begin{eqnarray}
    S_B = S_B(n_1,\ldots,n_B).
\end{eqnarray}
In fact, the details of the system outside the block are irrelevant. For example, 
different systems with different numbers of sites outside the block can have the
same $S_B$ as long as their $\{n_b\}$ are identical and the system is a 
Gaussian state. Thus to evaluate the function $S_B$, we can choose a simple 
system in which to evaluate it rather than using the actual system of interest.

Let's first consider a block with only one site ($B=1$). We would like 
to know the universal function $S_1(n_1)$. We choose a two site system 
containing a single particle, with normalized wavefunction
\begin{eqnarray}
    \ket{\psi} = (\sqrt{n_1}\hat{a}^{\dagger}_1+
            \sqrt{1-n_1}\hat{a}^{\dagger}_2)\ket{\Omega}.
\end{eqnarray}
The correlation matrix is
\begin{eqnarray}
    \begin{pmatrix} n_1 & \sqrt{n_1(1-n_1)} \\ \sqrt{n_1(1-n_1)} & 1-n_1 \end{pmatrix}
\end{eqnarray}
which has the required block correlation matrix $\Lambda_1=(n_1)$. The Schmidt 
decomposition of $\ket{\psi}$ is
\begin{eqnarray}
    \begin{aligned}
    \ket{\psi} &= \sqrt{1-n_1}(\ket{0})(\hat{a}^{\dagger}_2\ket{0})+
        \sqrt{n_1}(\hat{a}^{\dagger}_1\ket{0})(\ket{0}) \\
        &= \sqrt{1-n_1}\ket{0}\ket{1}+\sqrt{n_1}\ket{1}\ket{0}
    \end{aligned}
\end{eqnarray}
where $\ket{1}$ is the occupied state of the corresponding site. From this we see that
the reduced density matrix for site 1, $\hat{\rho}_1=\Tr_2[\ket{\psi}\bra{\psi}]$, is
\begin{eqnarray}
    \begin{aligned}
    \hat{\rho}_1 &= (1-n_1)\ket{0}\bra{0}+n_1\ket{1}\bra{1} \\
        &= (1-n_1)(\hat{I}_1-\hat{n}_1)+n_1\hat{n}_1
    \end{aligned}
    \label{rho1}
\end{eqnarray}
and
\begin{eqnarray}
    \begin{aligned}
    S_1(n_1) &= -\Tr[\hat{\rho}_1\log(\hat{\rho}_1)] \\
        &= -(1-n_1)\log(1-n_1)-n_1\log(n_1).
    \end{aligned}
\end{eqnarray}

For the system $\mathcal{B}$ with block size $B>1$, we can choose the system to be of 
size $2B$ and for each site in the block associate one site outside the block. The 
Gaussian state is the product state of the single particle states living on a pair, each 
identical in form to the $B=1$ state, with $B$ total particles. This system has no 
correlations or entanglement between these pairs. 
This means that the entanglement is the sum of the entanglement of each pair. Thus
\begin{eqnarray}
S_B(\{n_b\}) = \sum_{b\in\mathcal{B}}S_1(n_b)
\end{eqnarray}
which is identical to Eq.~\ref{entropy}. 
Note also that the overall reduced density matrix of the block is the product of the single 
site density matrices given in Eq.~\ref{rho1}.

An alternative argument can be made to derive the same equation which avoids the introduction
of a contrived environment. 
We could have taken as an ansatz that the reduced density matrix on $\mathcal{B}$, 
$\hat{\rho}_\mathcal{B}$, is the product of the single site reduced density matrix we 
derived in Eq.~\ref{rho1}, in other words 
$\hat{\rho}_\mathcal{B}=\otimes_{b\in\mathcal{B}}\hat{\rho}_b$.
We can show that this is in fact the unique reduced density of the state we are interested
in if we can show that it reproduces the correct correlation matrix of our state and is
a fermionic Gaussian state (that it obeys Wick's theorem). 
Both of these are easy to show explicitly.
Once we verify that this is indeed the correct reduced density matrix of our state, we
can calculate the entanglement entropy directly with 
$S_B=-\Tr[\hat{\rho}_{\mathcal{B}}\log(\hat{\rho}_{\mathcal{B}})]=-\sum_{b\in\mathcal{B}}
\Tr[\hat{\rho}_b\log(\hat{\rho}_b)]$, which matches Eq.~\ref{entropy}.

\section{GDMRG, an Algorithm to Obtain a Compressed Ground State Correlation Matrix as a 
GMPS}
\label{appendix2}

Here we describe fermionic Gaussian DMRG (GDMRG), a DMRG-like algorithm in the single 
particle context. 
The algorithm is an efficient method to directly obtain all the angles specifying
the compressed correlation matrix as a GMPS without ever needing to express the matrix in 
uncompressed form. The ground state GMPS of a hopping Hamiltonian on $N$ sites is calculated 
with a cost of only $O(B^3N)$, where $B$ is the block size of the GMPS (which determines the
accuracy of the compression and depends on the entanglement of the ground state).

\begin{figure}
    \subfloat[]{
      \includegraphics[width=0.45\textwidth]{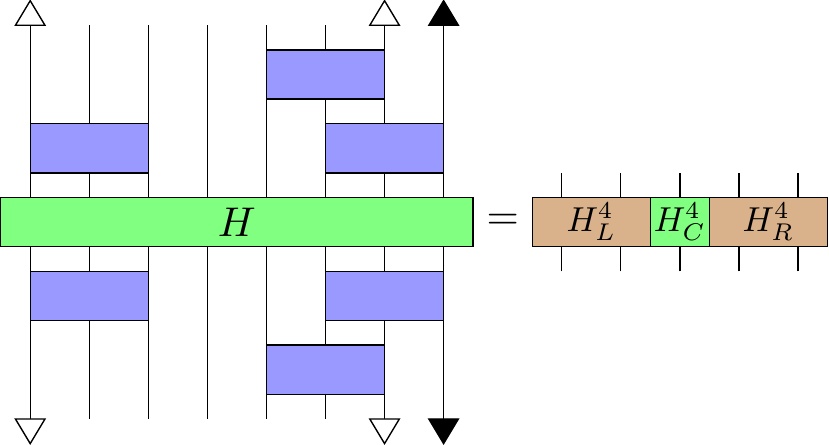}%
      } \ \ \ \ 
    \subfloat[]{
      \includegraphics[width=0.45\textwidth]{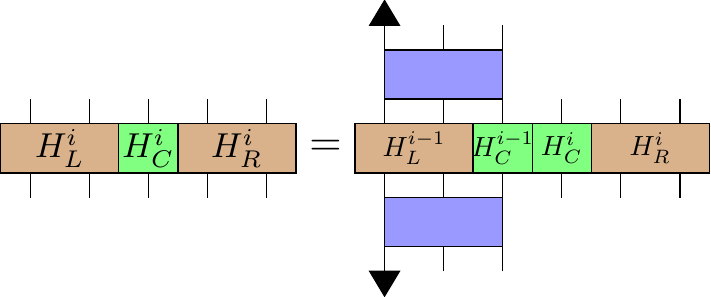}%
      }
\caption[]{ Fig.~\ref{gdmrg}(a) shows an example of an effective Hamiltonian centered at 
    site 4 for the GDMRG algorithm. The example is for $N=8$ sites and a 
    block size of $B=3$. Here the center is only one site, but could be more to improve 
    convergence just like in the standard DMRG algorithm. Fig.~\ref{gdmrg}(b) shows, for 
    a sweep to the right, how to obtain the new left block from the previous effective 
    Hamiltonian. }
\label{gdmrg}
\end{figure}

Imagine that we start with a hopping Hamiltonian $H$ on a lattice of $N$ sites, and we 
would like to obtain the GMPS with a block size $B$ that minimizes the energy of $H$. 
We begin with a random starting GMPS. Just like in the DMRG algorithm, we form an 
effective Hamiltonian centered at a site with a left and right block, which we show in 
Fig.~\ref{gdmrg}(a). Say that we start on the left side of the lattice and begin 
sweeping right. Our GMPS will start gauged to the left. For a single-site GDMRG, our 
center block is only one site, but we could use a larger center to decrease the number
of sweeps required for convergence. In practice for free gapless 
fermions we find that a single center site works quite well. The first step is to 
diagonalize the $2B-1$ site effective Hamiltonian, and obtain the effective 
correlation matrix $\Lambda_{\text{eff}}$. Using this $\Lambda_{\text{eff}}$, we 
diagonalize the first $B\times B$ block and, for a large enough $B$, find a fully 
occupied or unoccupied mode. Just as described in Fig.~\ref{mode}, we find the $B-1$ 
nearest neighbor $2\times 2$ gates that rotate $\Lambda_{\text{eff}}$ into the basis 
containing this mode, partially diagonalizing it. 
These gates form the first block of the GMPS. 

Next we would like to move the center to the right so that we can obtain the next block
of the GMPS. Because the compression is a unitary transformation, we can start moving the 
center to the right by undoing the gates in the block of the GMPS to the right of the center. 
This is step is in contrast to ordinary DMRG where a sequence of right blocks are stored and
are called from memory when needed.
We then obtain the effective Hamiltonian for the next site using the block of the GMPS we 
obtained from the previous effective Hamiltonian of the sweep, as shown in 
Fig.~\ref{gdmrg}(b). 
We repeat this process until we reach the end of the lattice, completing our first sweep. 
We continue sweeping back and forth until the energy is sufficiently converged. 
We use this algorithm to obtain a very accurate correlation matrix for systems up to 
$N=2^{16}=65536$, from which we obtain the GMPS in Fig.~\ref{block_scaling}. 
For $N=65536$ sites to obtain a correlation matrix with a relative error in the energy of 
less than $10^{-10}$, we require a block size of $B=22$ and 14 sweeps (where a single
sweep is from left to right or right to left).

\bibliography{gmps}

\end{document}